\documentclass[aps,pre,reprint,amsmath,amssymb,superscriptaddress,showpacs,floatfix]{revtex4-2}
\usepackage[english]{babel}
\usepackage{ragged2e}
\usepackage{graphicx}
\usepackage{dcolumn}
\usepackage{bm}
\usepackage[colorlinks, allcolors=blue]{hyperref}
\usepackage[usenames, dvipsnames]{color}
\usepackage{float}

\usepackage{lipsum}
\usepackage{titlesec}
\titlespacing\section{0pt}{12pt plus 4pt minus 4pt}{1pt plus 20pt minus 2pt}
\usepackage{xcolor}
\usepackage{tabularx}
\usepackage{amsmath}
\usepackage{comment}
\usepackage{afterpage}
\usepackage{placeins}
\usepackage{booktabs}
\usepackage{multirow}
\usepackage{array}
\usepackage{setspace}
\usepackage{siunitx}
\usepackage{hhline}
\usepackage{xfrac}
\usepackage{float,graphicx}
\usepackage{mathtools}
\usepackage{listings}
\usepackage{amssymb}
\usepackage[normalem]{ulem}
\usepackage{titlesec}
\usepackage{amsfonts}
\usepackage[version=4]{mhchem}

\usepackage{epstopdf}
\catcode`@11
\def\seceqaa{\@addtoreset{equation}{section}
principles\def\theequation{A\arabic{equation}}}
\def\seceqbb{\@addtoreset{equation}{section}
\def\theequation{B\arabic{equation}}}
\def\seceqcc{\@addtoreset{equation}{section}
\def\theequation{C\arabic{equation}}}
\def\seceqdd{\@addtoreset{equation}{section}
\def\theequation{D\arabic{equation}}}
\def\seceqee{\@addtoreset{equation}{section}
\def\theequation{E\arabic{equation}}}
\def\seceqff{\@addtoreset{equation}{section}
\def\theequation{F\arabic{equation}}}
\def\seceqgg{\@addtoreset{equation}{section}
\def\theequation{G\arabic{equation}}}
\def\seceqhh{\@addtoreset{equation}{section}
\def\theequation{H\arabic{equation}}}
\catcode`@11

\newcommand{\UGCDAE}{\affiliation{UGC-DAE Consortium for Scientific Research, Mumbai Centre, BARC Campus, Mumbai - 400 085, India}}

\newcommand{\PUNEUNIV}{\affiliation{Savithribai Phule Pune University, Ganeshkhind, Pune-411007, India}}

\newcommand{\IITB}{\affiliation{Department  of Physics, Indian Institute of Technology Bombay, Mumbai-400 076, India}}

\newcommand{\CRNTS}{\affiliation{Centre for Research in Nanotechnology and Science, Indian Institute of Technology Bombay, Mumbai-400 076, India}}

\begin{document}

\title{Intertwined Spin Chirality and Weyl Topology Driven Giant Hall Responses in Nd$_{4}$Sb$_3$}

\author{Anu}
\UGCDAE
\PUNEUNIV
\author{Subhasmita Kar}
\IITB
\author{Bhagyashree Pol}
\UGCDAE
\PUNEUNIV
\author{Debarghya Dutta}
\CRNTS

\author{Aftab Alam}
\email{aftab@iitb.ac.in}
\IITB
\CRNTS
\author{P D Babu}
\email{pdbabu@csr.res.in}
\UGCDAE

\begin{abstract}
Magnetic Weyl materials offer a unique platform for coupling magnetic order with real- and momentum-space topology, yet their simultaneous manifestation in a single material remains uncommon. Here, Nd$_4$Sb$_3$ is established as a highly spin-polarized magnetic Weyl semimetal exhibiting intertwined anomalous and topological Hall responses. Nd$_4$Sb$_3$ orders ferromagnetically below $T_{\mathrm{C}}\sim72$~K and undergoes spin reorientation near 42~K into a non-coplanar magnetic state. A large topological Hall resistivity of $\sim3.9~\mu\Omega$\,cm and anomalous Hall conductivity reaching $\sim1170$~S\,cm$^{-1}$ are observed. First-principles calculations identify an energetically preferred non-coplanar state with finite local scalar spin chirality, providing a microscopic route to the real-space Berry phase underlying the topological Hall effect. Calculated exchange interactions reveal dominant ferromagnetic exchange together with anisotropic exchange and finite Dzyaloshinskii--Moriya interactions, while Monte Carlo simulations yield a magnetic transition near 78~K, close to the experimental $T_{\mathrm{C}}$. Concurrently, magnetic reconstruction profoundly reshapes the reciprocal-space topology, increasing the number of Weyl-point pairs from 42 in the collinear ferromagnetic state to 115 in the non-coplanar state. Berry-curvature calculations yield an intrinsic anomalous Hall conductivity of $\sim983$~S\,cm$^{-1}$ at the experimentally determined carrier concentration, closely reproducing the experimental intrinsic contribution of $\sim990$~S\,cm$^{-1}$. These results establish Nd$_4$Sb$_3$ as a compelling platform where magnetic reconstruction intertwines real-space spin chirality with momentum-space Weyl topology.
\end{abstract}


\maketitle

\section{Introduction}

Magnetic Weyl semimetals and Weyl metals (MWSs) provide a particularly rich platform for exploring the interplay between magnetic order and nontrivial electronic topology. \cite{okamura2020giant, li2023emergence} In these systems, the breaking of time-reversal ($\mathcal{T}$) and/or inversion ($\mathcal{P}$) symmetry can lift band degeneracies and generate Weyl points, which act as sources and sinks of Berry curvature (BC) in momentum space. The distribution and energetic position of Weyl points are strongly influenced by the magnetic order, thereby providing an intrinsic route for controlling the Berry-curvature landscape. As a consequence, topological magnetic materials can exhibit pronounced intrinsic transverse responses, most notably the anomalous Hall effect (AHE) and anomalous Nernst effect, whose magnitudes are governed by the Berry curvature of the occupied electronic states. An additional degree of freedom emerges when Weyl topology is combined with half-metallicity, where the electronic states at the Fermi level are strongly spin polarized. The simultaneous presence of magnetic order, spin-polarized electronic states, and tunable Weyl topology therefore offers an attractive platform for realizing large Berry-curvature-driven transport responses and for exploring their potential in spintronic applications. \cite{hu2015chern, zhang2021weyl}

The Berry-curvature-driven AHE has been extensively investigated in a variety of magnetic topological materials, encompassing both ferromagnetic (FM) and antiferromagnetic (AFM) systems, including GdAlSi\cite{nag2024gdalsi}, Mn$_3$Sn\cite{li2023field}, EuAgAs\cite{jin2021multiple} and Co$_2$MnGa\cite{leiva2021giant}. In particular, rare-earth-based noncentrosymmetric magnetic materials have emerged as an important class in which magnetic order and crystal symmetry can provide complementary control over the topological electronic structure. Weyl semimetallic behavior has recently been reported in several members of the RAlSi/Ge family (R: rare earth),\cite{kunze2024optical} as well as in NdAlSi \cite{wang2022ndalsi} and CeAlSi\cite{cheng2024tunable}. These developments raise an important question: can other rare-earth compounds host a combination of strong spin polarization, Weyl topology, and an enhanced intrinsic Hall response?

Recently, some of the R$_4$Sb$_3$ type of compounds with non-centrosymmetric cubic anti Th$_3$P$_4$ type crystal structure were found to display topological properties. One such compound, Gd$_4$Sb$_3$, which was theoretically predicted to show half metalicity and interesting topological features such as anomalous Hall and Nernst conductivities,\cite{id10} was experimentally verified \cite{id9} along with band structure calculations. The large anomalous Nernest effect observed in  Gd$_4$Sb$_3$ was reported to be arising from the intrinsic Berry curvature near E$_F$ induced by topological non trivial energy band structure.\cite{id9}
On the other hand, the isostructural Ce$_4$Sb$_3$ compound exhibits altermagnetism as well as Dirac fermions and serves as a promising platform for studying the interplay between topology and altermagnetism\cite{id8}.  These observations prompted us to investigate in detail another isostructural compound Nd$_4$Sb$_3$ on which there are very few experimental reports of magnetic and transport properties. \cite{id7, R4X32011IC} Based neutron diffraction and magnetization measurements Nd$_4$Sb$_3$ is reported to exhibit collinear  ferromagnetism  below 77 K and undergoes non-collinear ferromagnetic phase below 38 K. However, these reports suffer from substantial impurity phase of NdSb up to 17$\%$, \cite{id7, R4X32011IC}.
In this context, Nd$_4$Sb$_3$ provides a compelling yet unexplored platform. Despite the growing interest in rare-earth-based topological materials, its electronic structure, magnetic topology, and Berry-curvature-driven transport properties have not, to our knowledge, been systematically investigated. Establishing the relationship between its magnetic electronic structure, Weyl-point distribution, and anomalous Hall response can therefore provide insight into how magnetic order and noncentrosymmetric crystal symmetry cooperate to generate large intrinsic transverse transport. In this work, we investigate the electronic and topological properties of Nd$_4$Sb$_3$ and demonstrate a strongly spin-polarized electronic structure together with a dense distribution of Weyl points in the vicinity of the Fermi level. We further establish the resulting Berry-curvature enhancement of topological Hall response and its connection to the large anomalous Hall conductivity, thereby identifying Nd$_4$Sb$_3$ as a promising platform for studying the interplay between spin polarization, Weyl topology, and intrinsic anomalous Hall transport.

\section{Experimental Details}

Polycrystalline compound of Nd$_4$Sb$_3$ was synthesized by using arc melting in argon atmosphere by taking high purity starting materials  Nd (99.9$\%$) and Sb (99.99$\%$) in stoichiometry ratio.To ensure the homogeneity of the sample, the ingot was remelted several times by turning the sample upside down. Optimization of synthesis was tried with different amounts of Sb excess. The final compound, obtained with 5$\%$ Sb excess, found to have a weight loss of less than 0.1$\%$. The powder X-Ray diffraction (XRD) pattern was recorded with the help of Bruker D2 Phaser XRD machine using a wavelength (Cu -  $K_\alpha$) 1.5406 $\text{\AA}$. The temperature and magnetic field dependent magnetization measurements were carried out using Vibrating sample magnetometer (VSM) probe attached with the commercial Quantum Design Physical Property Measurement System (PPMS). The Heat capacity was measured from 2 K to 310 K under zero and different fields using Heat capacity option of PPMS, which employs 2-tau  relaxation calorimetry. Electrical transport experiments including Hall effect as function of temperature and magnetic field were performed using electrical transport option (ETO) of the PPMS, which uses low frequency ac excitation.  Electrical contacts on bar shaped sample were made using silver paste.

\section{Computational Details}

First-principles calculations were performed within density functional theory (DFT) using the Vienna \textit{Ab initio} Simulation Package (VASP) \cite{kresse1993ab, kresse1996efficiency} and the projector augmented-wave (PAW) method. \cite{kresse1999ultrasoft} The exchange-correlation interaction was treated within the generalized gradient approximation (GGA) using the Perdew--Burke--Ernzerhof (PBE) functional. \cite{perdew1997generalized} The Brillouin zone (BZ) was sampled using a $\Gamma$-centered $6\times6\times6$ $k$-point mesh, and a plane-wave kinetic-energy cutoff of 380 $eV$ was employed. The total energy convergence criterion for the self-consistent calculations was set to $10^{-6}$ $eV$. Calculations were performed both in the absence and presence of spin-orbit coupling (SOC). To account for the localized nature of the Nd-$4f$ electrons, an effective Hubbard parameter of $U=6.0$ eV was applied to the Nd sites within the Dudarev formalism. \cite{dudarev1998electron}

To determine the interatomic exchange parameters, $J_{ij}$, from first-principles electronic-structure calculations, we employed the Green's-function formalism introduced by Liechtenstein {\it et al.}, commonly referred to as the Liechtenstein--Katsnelson--Antropov--Gubanov (LKAG) method. \cite{liechtenstein1987local} This approach provides an efficient framework for evaluating the change in total energy, $\delta E$, associated with infinitesimal rotations of local magnetic moments.  A key advantage of the LKAG approach is that the exchange interactions can be evaluated directly from the ground-state electronic structure of the primitive unit cell, eliminating the need to construct computationally expensive large supercells or perform extensive total-energy mapping calculations. The method requires the electronic Green's function corresponding to an effective tight-binding Hamiltonian. In practice, this Hamiltonian is commonly constructed using maximally localized Wannier functions (MLWFs), typically generated with the WANNIER90 package. \cite{marzari2012maximally, souza2001maximally} Further, Monte Carlo simulations were performed to explore the finite-temperature magnetic response based on the following Hamiltonian,
\begin{equation}
    H = -\sum_{i,j(i\neq j)} \left[\boldsymbol{J_{ij}^\mathrm{iso}} \hat S_i. \hat S_j +\hat{S_i} \boldsymbol{J_{ij}^\mathrm{ani}} \hat{S_j} +\boldsymbol{D_{ij}} \cdot (\hat S_i\times \hat S_j )\right]
     \label{hamilton}
\end{equation}
which contains the isotropic exchange contribution $J_{ij}^\mathrm{iso}$, the symmetric anisotropic exchange contribution $J_{ij}^\mathrm{ani}$, and the antisymmetric Dzyaloshinskii–-Moriya interaction (DMI), represented by the vector $\boldsymbol{D_{ij}}$. $\hat{S_i}$ is a $3\times3\times1$ unit vector oriented along the magnetic moment $m_i$ at site $i$. The isotropic term is independent of the orientation of the magnetic pair relative to the crystal lattice, whereas the anisotropic and antisymmetric terms describe orientation-dependent exchange and DMI, respectively.
At each temperature, 100000 steps were discarded for equilibration, followed by 100000 sampling steps. The magnetic transition temperature, corresponding to the Curie temperature in the ferromagnetic case, was identified from the slope change of magnetization (M) / susceptibility ($\chi$) vs temperature (T) data, calculated by the following equations:
\begin{equation}
  M = \frac{1}{N} \sum_{i = 1}^N S_{i}
   \label{eq:nag}
\end{equation}
\begin{equation}
\chi = \frac{\langle M^2\rangle - \langle M\rangle^2}{k_\mathrm{B} T},
\label{eq:Cm}
\end{equation}
\textcolor{black}{where, $k_\mathrm{B}$ is the Boltzmann constant, T, N and $\langle$M$\rangle$ are the temperature, number of magnetic atoms in the supercell and thermal average of magnetization respectively.}

To investigate the topological nature of the electronic structure and surface states, a MLWF Hamiltonian was constructed from the GGA+$U$+SOC Bloch states using the WANNIER90 package. \cite{marzari2012maximally, souza2001maximally} A total of 360 bands were included in the Wannierization, with projections onto the Nd ($s$, $p$, $d$, and $f$) and Sb ($s$ and $p$) orbitals. The resulting Wannier Hamiltonian was subsequently employed to obtain the surface spectral function and Fermi surface of the (001) surface using the iterative Green's-function method \cite{lee1981simple, sancho1985highly} for a semi-infinite slab, as implemented in the WannierTools package. \cite{wu2018wanniertools} The anomalous Hall conductivity (AHC) was evaluated within the intrinsic Berry-curvature framework using the postw90 module \cite{pizzi2020wannier90} of the Wannier90 package, with Wannier interpolation used to efficiently integrate the Berry curvature throughout the BZ.
The carrier concentrations were obtained from the total density of states (DOS) calculated within DFT. The electron and hole concentrations at a given chemical potential $\mu$ were evaluated on an energy grid containing 8001 points according to
\begin{equation}
n_e(\mu) = \int_{E_C}^{E_{\mathrm{max}}} g(E) \, f(E,\mu) \,dE,
\end{equation}
and
\begin{equation}
n_h(\mu) = \int_{E_{\mathrm{min}}}^{E_V} g(E) \left[1-f(E,\mu)\right] \, dE.
\end{equation}
where $n_e$ and $n_h$ denote the electron- and hole-type carrier concentrations, respectively, while $g(E)$ is the total DOS and $f(E,\mu)$ is the Fermi--Dirac distribution at energy $E$ and chemical potential $\mu$. $E_{\mathrm{min}}$ and $E_{\mathrm{max}}$ define the lower and upper bounds of the energy window used for the integration, whereas $E_V$ and $E_C$ denote the valence- and conduction-band edges, respectively. In the present analysis, both $E_V$ and $E_C$ are taken at the Fermi level, $E_F$.

\begin{figure}[t]
\centering
\includegraphics[width=0.48\textwidth]{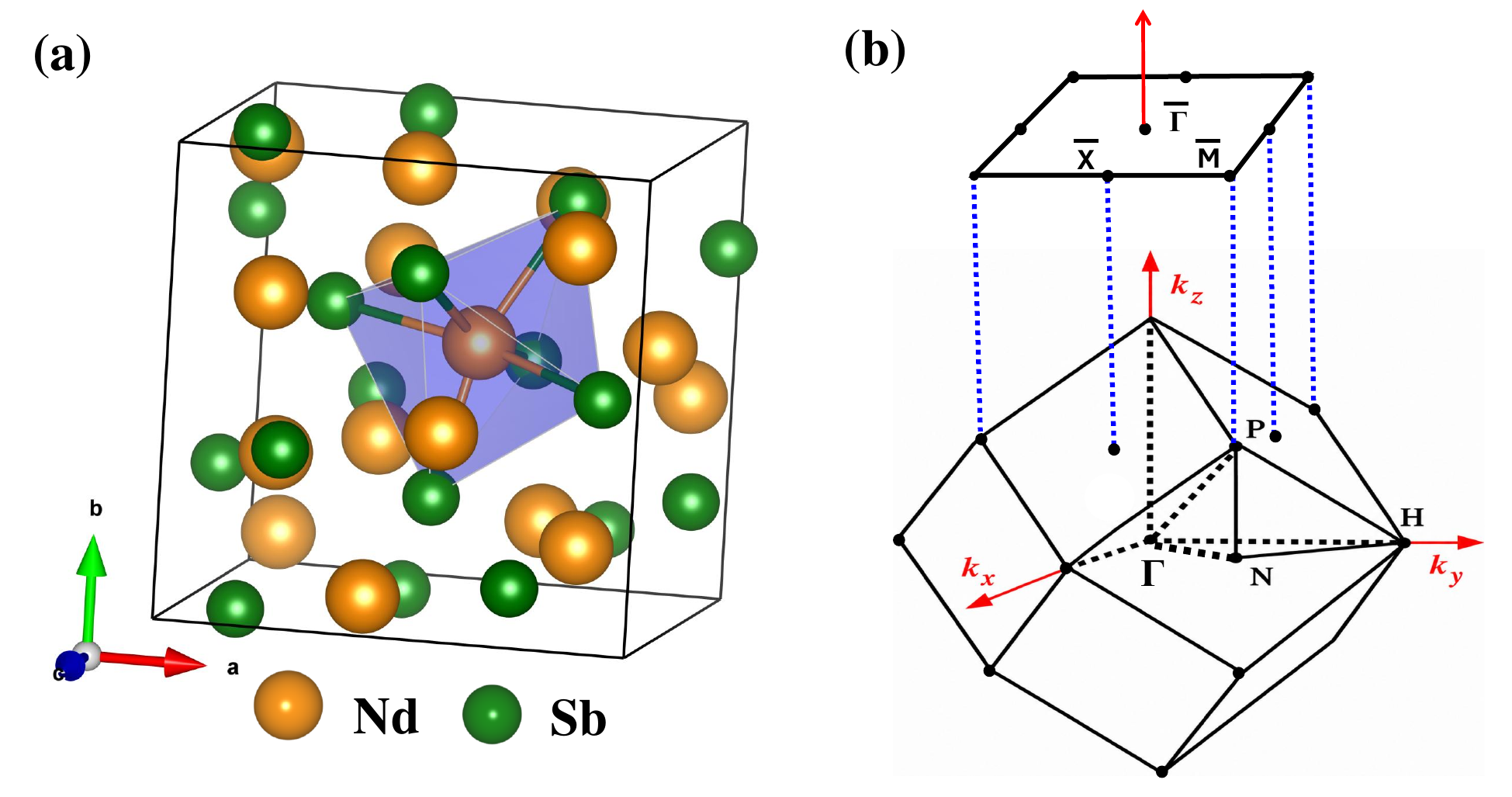}
\caption{(a) Crystal structure of Nd$_4$Sb$_3$, and (b) its bulk and surface Brillouin Zone (BZ) with high symmetry points.}
\label{cstr}
\end{figure}

\begin{figure*}[t]
\centering
\includegraphics[width=0.95\linewidth]{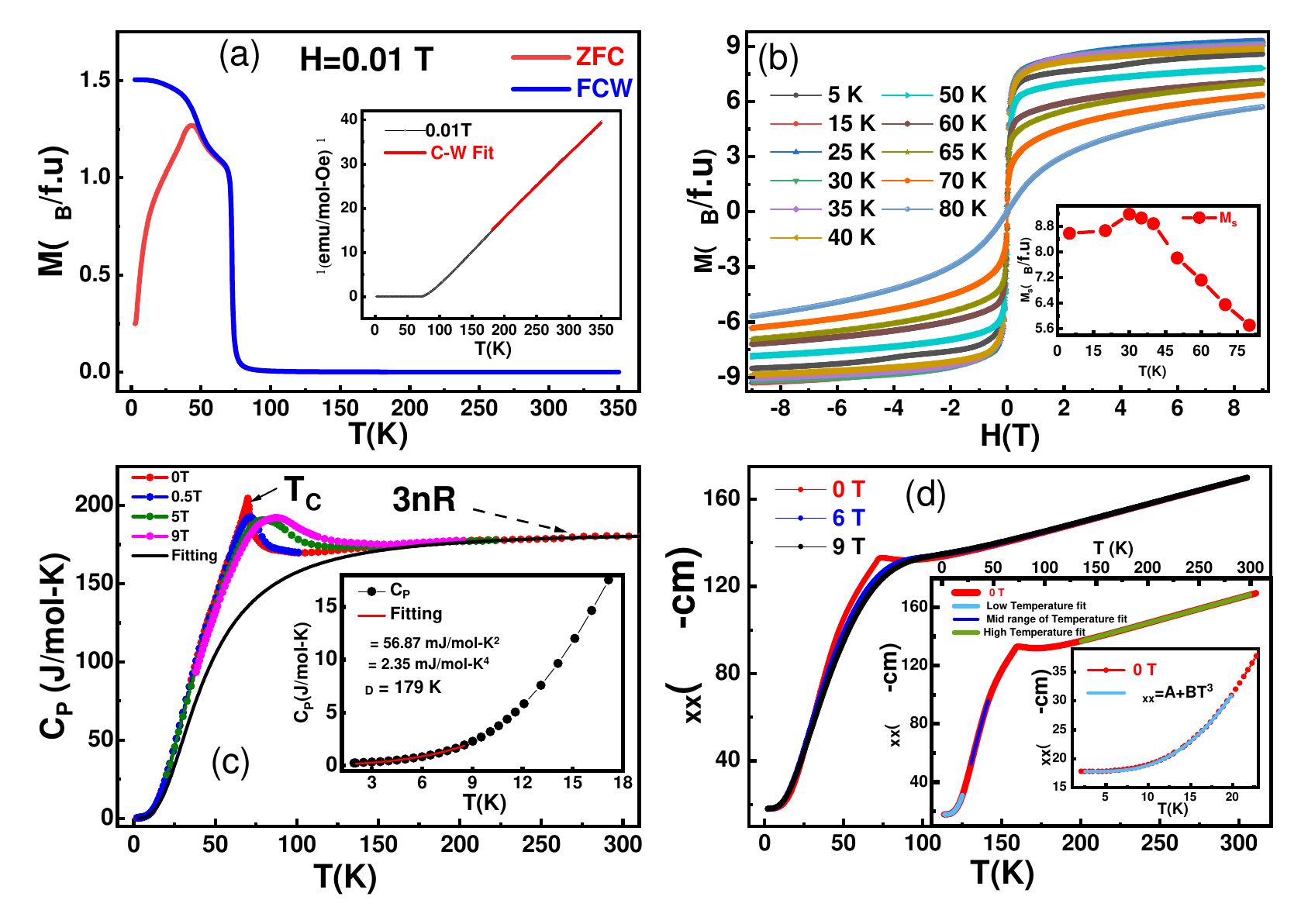}
\caption{(a) ZFC and FCW magnetization curves for Nd$_{4}$Sb$_{3}$; (Inset) the CW fit to $\chi^{-1}$ data in paramagnetic region. (b) Magnetization versus field at selected temperatures; (Inset) Saturation magnetization at 9 T vs. temperature. (c) Temperature dependence of heat capacity under different fields along with simulated (black) curve for lattice and electronic parts; (Inset) Fit to C$_{\text{P}}$ = $\gamma$ T+$\beta$T$^3$ for $T <$ 10~K.(d) $\rho_{xx}$ vs. T under  different fields; (Inset) various fits  to $\rho_{xx}$ data (see text for more details).} \label{mhcrt}
\end{figure*}

\section{Results and discussion}

Rietveld refinement of X-ray diffraction reveals that the compound Nd$_4$Sb$_3$ crystallizes in anti-Th$_3$P$_4$ type cubic  structure with $I\bar{4}3d$ space group. In this structure, Nd and Sb atoms occupy 16$c$ and 12$a$ Wyckoff sites, respectively. The crystal structure is shown in Figure \ref{cstr}(a) and the refined XRD pattern is shown in Figure S1 of supporting information. The lattice parameters deduced from the refinement are $a = b = c =  9.4053(4)$ $\text{\AA}$, which are in good agreement with the previous report.\cite{id7}

\subsection{Magnetic Characterization}
The dc magnetization data recorded in the `Zero-Field Cooled' (ZFC) and `Field Cooled' (FC) protocols in the field of $H$ = 0.01 T clearly show the compound orders ferromagnetically at $T_C$ $\sim$ 72 K, as shown in Figure \ref{mhcrt}(a). This was followed by a spin-reorientation transition at $T_{SR}$ $\sim$ 42 K possibly to an antiferromagnetic / canted spin structure, consistent with previously reported observations \cite{id7}. Below 42 K, the ZFC and FC bifurcation is very large; the FC curve continues to rise before leveling off as $T$ approaches 2 K. On the other hand, the ZFC curve continues to decrease and tends to level off at a finite value of M below 2.7~K.  This indicates a finite ferromagnetic moment even at 2~K and points to a complex spin structure below 42 K. Susceptibility $\chi$ ($= M/H$) data in the paramagnetic region was analyzed in terms of Curie-Weiss (CW) law ($\chi$ = $C/$($T - \theta _p$) as shown in the inset of Figure \ref{mhcrt}(a). CW analysis yields a value of $\theta_p$  = 74.6 K, which matches well with the previous report \cite{id7}. The positive value of $\theta_p$ indicates the ferromagnetic nature of the compound below $T_C$.  The effective moment, $\mu_{eff}$, deduced from the Curie constant using the relation $\mu_{eff}$ = 2.828$\sqrt{CA}$ (where A - molecular weight), turns out to be 3.58 $\mu_B$ /Nd$^{3+}$ atom which is quite close to the theoretical estimate of 3.62 $\mu_B$ per Nd$^{3+}$ atom.
The isothermal magnetization curves $M(H)$ as a function of the magnetic field at several temperatures are  shown in Figure \ref{mhcrt}(b). These MH curves are typical of soft ferromagnetic systems in the entire temperature range with negligible coercivity. However, MH curves below 30~K show small coercivity, its value being 15~mT at 5~K,  which gradually decreases to become insignificant by $\sim$30~K. The variation of saturation magnetization ($M_S$, the value of M at 9~T) is shown in the  inset of Figure \ref{mhcrt}(b). The maximum value of $M_S$ is observed at $\sim 30$~K and decreases with increasing $T$. For $T <$ 30 K, the $M_S$ values are slightly lower suggesting that the spin canting is more at low temperatures. The corresponding saturation magnetic moment at 5 K is $\sim$2.14 $\mu_B$/Nd atom, while the maximum value is $\sim$2.3 $\mu_B$/Nd atom at 30 K. This, together with the drop in ZFC below 42 K, is suggestive of a non-collinear or canted  spin structure at low temperatures, and even 9 T magnetic field is unable to align all the moments in the field direction.

A large bifurcation between ZFC and FC curves, the drop in ZFC could also point to the presence of  spin-glass-like state at low temperatures. To investigate this aspect, detailed ac susceptibility ($\chi_{ac}$) measurements were performed on this compound as a function of temperature over a wide frequency range from 97~Hz to 9717~Hz. The temperature dependence of real ($\chi^{\prime}$) and imaginary ($\chi^{\prime\prime}$) parts of ac susceptibilities, (shown in Figure S2 of the supporting information) exhibits two  peaks,  one sharp peak around 72 K corresponding to $T_C$ and another one around $T_{SR}$ $\sim$ 42 K consistent with the transitions seen in dc magnetization. All the curves corresponding to different frequencies coincide with each other around $T_C$ but a certain amount of frequency ($f$) dispersion is seen around 42 K and below. This peak around 42 K shifts slightly towards high temperatures with increasing $f$ which is a typical signature of spin-glass systems. However, this shift appears to be small and most likely indicates a weak canonical spin-glass-like behavior. Since this frequency dispersion is on the rising curve of $\chi_{ac}^{\prime}$, it is not feasible to deconvolute the peak and analyze $T_f$ vs $f$ data in terms of various spin-glass models.

\subsection{Heat Capacity}

Figure \ref{mhcrt}(c) shows the temperature dependence of specific heat ($C_p$). A sharp peak corresponding to $T_C$ was observed around 71 K but no visible feature around $T_{SR}$. With an increasing magnetic field, the peak at 71 K broadens, shifts to the right and is progressively suppressed. Even 9~T magnetic field is not sufficient enough to completely suppress this broad peak. This shift of the peak to the right indicates that the magnetic order of the compound below 71 K is ferromagnetic. The saturation value of $C_p = 176$ JK$^{-1}$mol$^{-1}$ near room temperature is in good agreement with the Dulong-petit limit $C_p$ = 3nR = 174.4~JK$^{-1}$mol$^{-1}$, where n = 7 is the number of atoms per formula unit, R is the ideal gas constant. A linear fit to the low temperature $C_P$ data using the relation $C_P$ = $\gamma T$ + $\beta$T$^3$ yields $\gamma$ = 56.87 mJK$^{-2}$mol$^{-1}$ and $\beta$ = 2.35 mJK$^{-4}$mol$^{-1}$. The Debye temperature ($\theta_D$) is then determined to be 179~K using the expression $\theta_D = \sqrt[3]{12\pi ^4 N K_{B}/(5\beta)}$. The density of state at Fermi level (E$_f$), obtained with the help of the relation $n(E_f)$ = 3$\gamma$/($\pi K_B$)$^2$ is 24 states/eV-f.u. To extract the magnetic part of $C_P$, the lattice and electronic parts of $C_P$ must be estimated and subtracted from the total $C_P$. The lattice and electronic contributions are simulated using the expression
\[
C_p(T) = \gamma T + 9nR \left(\frac{T}{\Theta_D}\right)^3 \int_0^{\Theta_D/T}
\frac{x^4 e^x}{(e^x-1)^2}\,dx \, ,
\]
which is shown as a black line in Figure \ref{mhcrt}(c). The simulated curve was then subtracted from the total $C_P$ to obtain the magnetic part, $C_{mag}$; subsequently, the magnetic entropy ($S_m$) was calculated using the relation $S_m$ = $\int  dT \, C_{mag}/T $ (shown in Figure S3 of the supporting information). The value $S_m$ $\sim$ 16.6 JK$^{-1}$mol$^{-1}$/$Nd^{3+}$ at the transition is lower than 19.14 JK$^{-1}$mol$^{-1}$/$Nd^{3+}$, the theoretical estimate of Rln10 ($\equiv$ Rln(2J+1) with J = 9/2 for $Nd^{3+}$) for Nd atom. This indicates that not all Nd spins are participating in the transition.

\subsection{Longitudinal Electrical Transport}
\begin{figure*}
    \centering
\begin{minipage}[b]{0.48\textwidth}
    \centering
    \includegraphics[width=\linewidth]{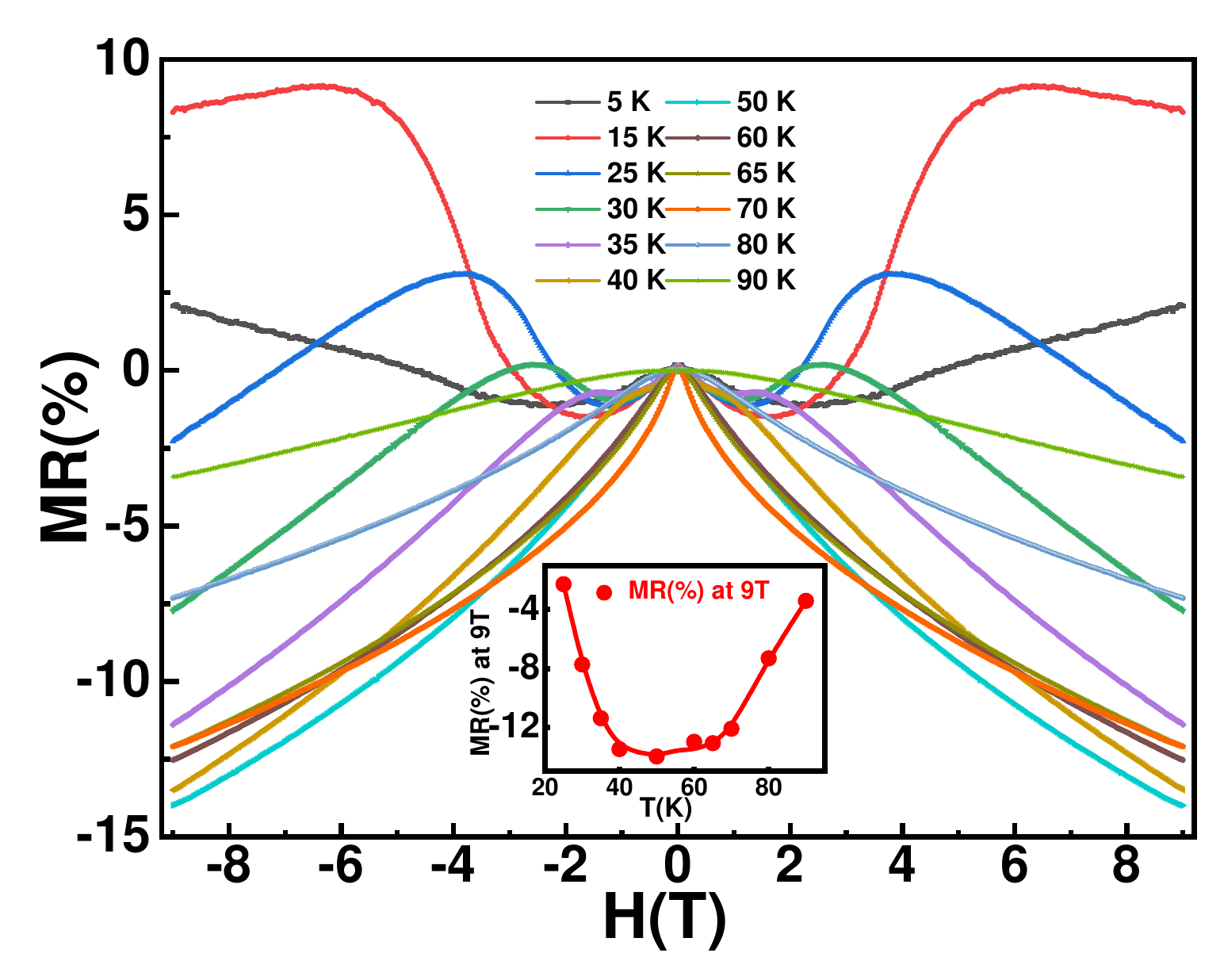}
    \caption{Magnetoresistance (MR) vs. field (H) at several temperatures; (Inset) 9~T-MR vs. T.}
    \label{MR}
\end{minipage}
\hspace{0.5cm}
\begin{minipage}[b]{0.48\textwidth}
    \centering
    \includegraphics[width=\linewidth]{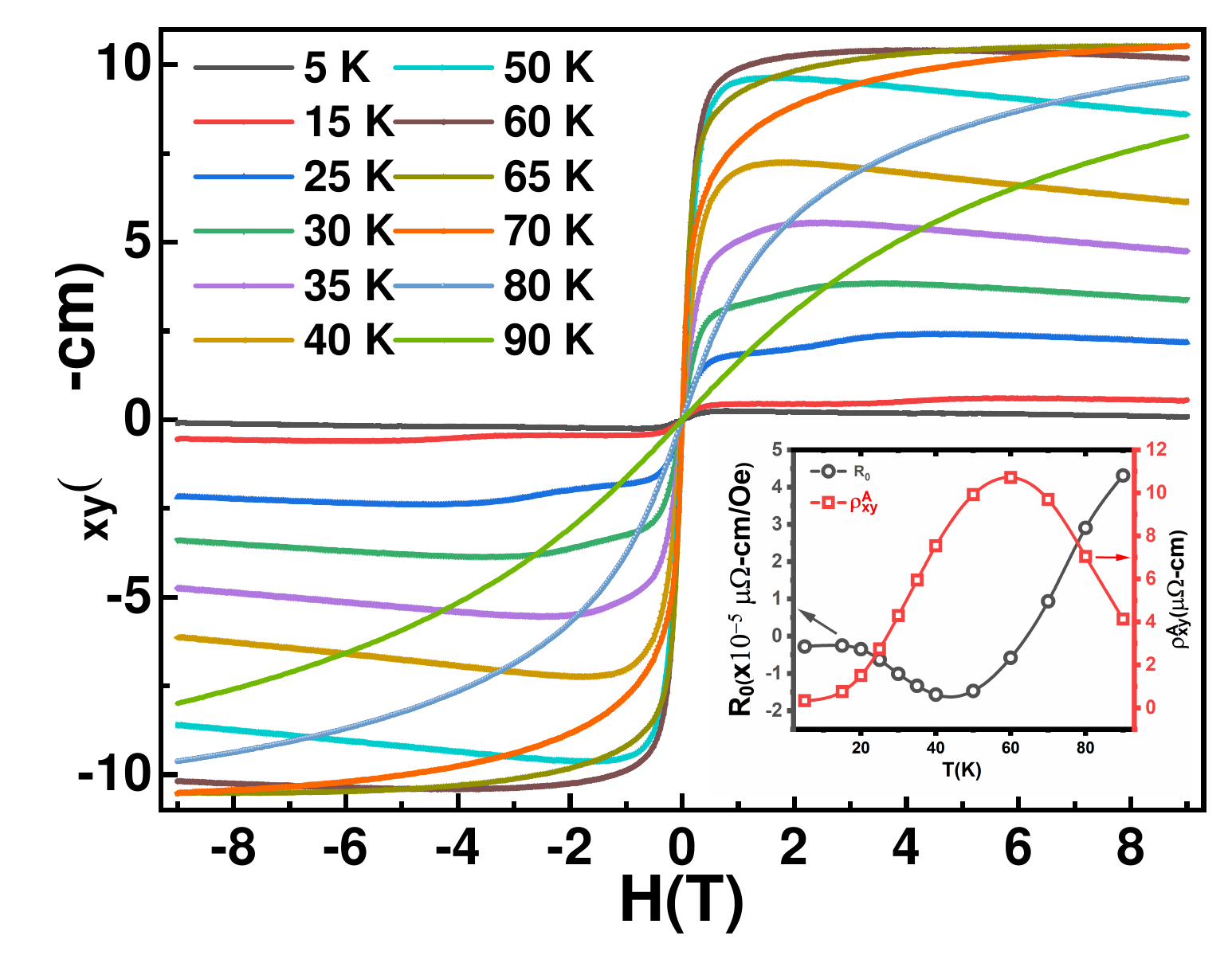}
    \caption{Symmetrized total Hall resistivity ($\rho_{xy}$) vs. field (H); (Inset) variation of ordinary Hall coefficient (R$_0$) and anomalous Hall resistivity ($\rho_{xy}^{A}$) with T.}
    \label{HR}
\end{minipage}
\end{figure*}

The temperature dependence of longitudinal resistivity ($\rho_{xx}$), displayed in Figure \ref{mhcrt}(d), is typical of metallic systems. A sharp slope change, almost like a peak, is observed around $T_C$. The resistivity of Nd$_{4}$Sb$_{3}$ is in $\mu \Omega$- cm range (varying from  17.6 $\mu \Omega$- cm at 5 K and 168.6 $\mu \Omega$-cm at 300 K), indicating its metallic nature. The residual resistivity ratio (RRR), $\rho_{300 K}/\rho_{5 K}$, is 9.6, which is comparable to that of heavy rare-earth based compound GdCo$_2$.\cite{qiu2025coexistence} The application of a magnetic field reduces the sharpness of peak-like feature seen around $T_C$, turning into a broad, rounded curve. According to the Matthiessen's rule, the total resistivity of the alloy is the sum of all contributions coming from the various scattering mechanisms and can be expressed as  $\rho_{xx}$(T) = $\rho_0$ + $\rho_{e-m}$ + $\rho_{e-p}$ + $\rho_{e-e}$, where $\rho_{e-m}$, $\rho_{e-p}$ and $\rho_{e-e}$ represent the resistivity contributions coming from electron-magnon, electron-phonon  and electron-electron scattering mechanisms, respectively.\cite{bid2006temperature} Here, $\rho_0$ is the temperature independent part caused by defects, impurities, and disorder.\cite{belopolski2019discovery} However, these scattering mechanisms have different functional dependencies on T in different temperature regimes. Hence, resistivity data for $T \leq$ 20~K was fitted (inset of Figure \ref{mhcrt}(d)) to a power law as given by
\begin{equation}
\rho_{xx} = \rho_0 + A T^n
\end{equation}
yielding an exponent value of $n$ = 3. Furukuwa, \cite{furukawa2000unconventional} taking into account the non-rigid band behavior arsing from spin fluctuations at finite temperatures in half metallic systems, showed that  $\rho_{xx}$  varies as  (T/D$_s$)$^3$  (where D$_s$  is the spin wave stiffness coefficient) coming from unconventional one magnon scattering at low temperatures. This $T^3$ variation of $\rho_{xx}$ has been previously observed in several half-metallic compounds such as Fe$_2$TiSn,\cite{chaudhuri2021half} Fe$_2$RhSi,\cite{FeRhSi2021KGS} CeAuSb$_2$ \cite{balicas2005magnetic} and Co$_2$CrAl.\cite{zhang2004heusler} Therefore, the presently observed $T^3$ resistivity variation provides clear experimental evidence for the half metallic nature of Nd$_4$Sb$_3$. In the intermediate temperature range of 28 K to 45 K, the resistivity is well described by the $\sqrt{T}$ term (inset of Figure \ref{mhcrt}(d)) of the form
 \begin{equation}
\rho_{xx} = \rho_0 + A \sqrt{T} \, ,
\end{equation}
which arises from the $e-e$ scattering. Well above $T_C$, in the high-temperature region (135 K to 300 K), the temperature variation of $\rho_{xx}$ is well accounted for by a linear T term that arises mainly from electron phonon scattering as in normal metals.
The magnetoresistance (MR), defined  as MR(\%) =  100$\times$$({\rho_{xx}(H)  - \rho_{0} })/{\rho_{0}}$, as a function of magnetic field for different temperatures, is shown in Figure \ref{MR}. The behavior of MR below and above $T_{SR}$ is quite different. Above 40 K, the MR is completely negative with a maximum value of approximately 15\% for H = 9 T at 50 K. Above $T_C$, the variation of MR with H is not that significant, although it remains negative.  Below 40 K, MR is $\omega$-shaped up to 4--5~T. Initially negative, the MR changes sign to become positive and goes through a broad maximum at a certain critical field before showing a slow decrease. This trend, the fields at which MR changes sign and reaches its maximum, etc., is sensitive to temperature. While a maximum positive MR of $\sim$9\% is observed at 6 T for T = 15~K, it drops to $\sim$3\% at 4 T for 25~K. As T increases toward 40 K, the critical field and the peak MR values decrease. At 5~K, the critical field seems to be larger than 9~T; therefore, neither the peak nor the downward trend is observed. The absolute value of negative MR at 9 T increases from 25 K to 50 K and then begins to decrease as T goes beyond $T_C$ as shown in the inset of Figure \ref{MR}. The high sensitivity of MR below 40~K to temperature and magnetic field reflects field- and temperature-induced spin dynamics arising from non-collinear spin structure. \cite{id13, li2020giant} This experimental observation is consistent with non-coplanar spin structure predicted by theoretical simulations described below. Sign change positive to negative MR was also observed in Fe$_3$Sn$_2$ \cite{li2019large} and Fe$_5$Sn$_3$. \cite{li2020large}

\subsection{\textbf{Hall Effect}}

The Hall resistivity ($\rho_{xy}$) of Nd$_{4}$Sb$_{3}$ measured in both positive and negative fields was anti symmetrized using the relation $\rho_{xy} = ({\rho_{xy}(+H) - \rho_{xy}(-H)})/{2}$, where ``+'' and ``-'' signs refer to positive and negative applied magnetic fields.

\begin{figure*}[t]
\centering
\includegraphics[width=0.98\linewidth]{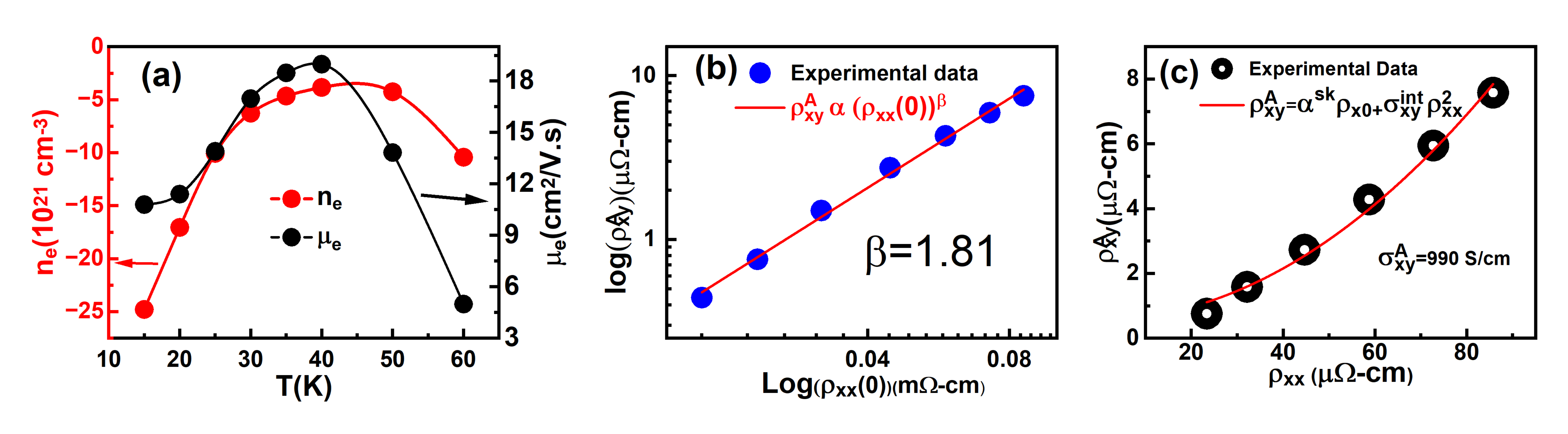}
\caption{(a) Carrier concentration (n$_0$) and mobility ($\mu_e$) vs. T (b) Variation of log($\rho_{xy}^A$)  vs. log($\rho_{xx}$)  and the corresponding linear fitting, and (c) $\rho_{xy}^A$ vs. $\rho_{xx}$, solid line is the fit as per the TYJ scaling (see text for more details).} \label{nmu-scal}
\end{figure*}

Figure \ref{HR} shows the variation of symmetrized hall resistivity  ($\rho_{xy}$) data as a function of field at different temperatures. The  variation of $\rho_{xy}$ is non linear at low fields, quickly saturates, and varies linearly at high fields. The Hall effect in a ferromagnet consists of two parts: the ordinary Hall effect (OHE) ($\rho_{xy}^{N}$) and the anomalous Hall effect (AHE) ($\rho_{xy}^{A}$). The OHE arises from the Lorenz force and varies linearly with field as  $\rho_{xy}^{N}$ = $R_{0}$H, where $R_0$ is the  ordinary Hall coefficient. On the other hand, AHE in magnetic materials arises from intrinsic (Berry curvature and side-jump scattering due to  spin-orbit interaction) and extrinsic (skew scattering) mechanisms.\cite{karplus1954hall} To estimate each of these contributions, the linear part of $\rho_{xy}$ above 4 T was fitted to the relation $\rho_{xy}$ = $\rho_{xy}^N$ + $\rho_{xy}^A$ = R$_0 H$ + $\rho_{xy}^A$, which yields the anomalous part as the intercept and the ordinary Hall coefficient, R$_0$, from the slope.  The deduced values of R$_0$ are plotted against T in the inset of Figure \ref{HR}. R$_0$ is negative for $T <$ 65~K and becomes positive for $T >$ 65~K, indicating that the dominant charge carriers switch from electrons to holes around 65~K. Similar behavior was reported for Fe$_5$Sn$_3$ \cite{li2020large} and Mn$_2$PtSn.\cite{liu2018giant} The anomalous Hall resistivity ($\rho_{xy}^A$) exhibits a broad peak-like variation with temperature, as displayed in the inset of Figure \ref{HR}. The maximum is around 60~K with a peak value of $\rho_{xy}^A$ is 10.7 $\mu \Omega$-cm around 60 K; its value drops on either side of 60~K. The carrier concentration ($n_e$) and carrier mobility ($\mu_e$) are determined using the single-carrier Drude Model relations, $n_e$ = 1/$R_0 e$ and $\mu_e$ = $1/ne\rho_{xx}$(0). Figure \ref{nmu-scal}(a) shows the variations of  $n_e$ and $\mu_e$ with temperature. The value of $n_e$ decreases from $\sim$ -2.5$\times$10$^{22}$ cm$^{-3}$ at 15 K to -3.84$\times$10$^{21}$ cm$^{-3}$ around 50 K, beyond which it increases. The carrier mobility, $\mu_e$, on the other hand, increases from its values around 15 K, exhibiting a maximum at 40  K with a value of 19 cm$^2$/V-s before dropping sharply with a further increase in temperature. Such variations in $n_e$ and $\mu_e$ along with temperature induced change in dominant carriers type have also been reported earlier in Mn$_2$FeSn,\cite{liu2022large} Mn$_2$PtSn \cite{liu2018giant} and Fe$_5$Sn$_3$. \cite{li2020large}

\begin{figure*}[t]
\centering
\includegraphics[width=0.98\linewidth]{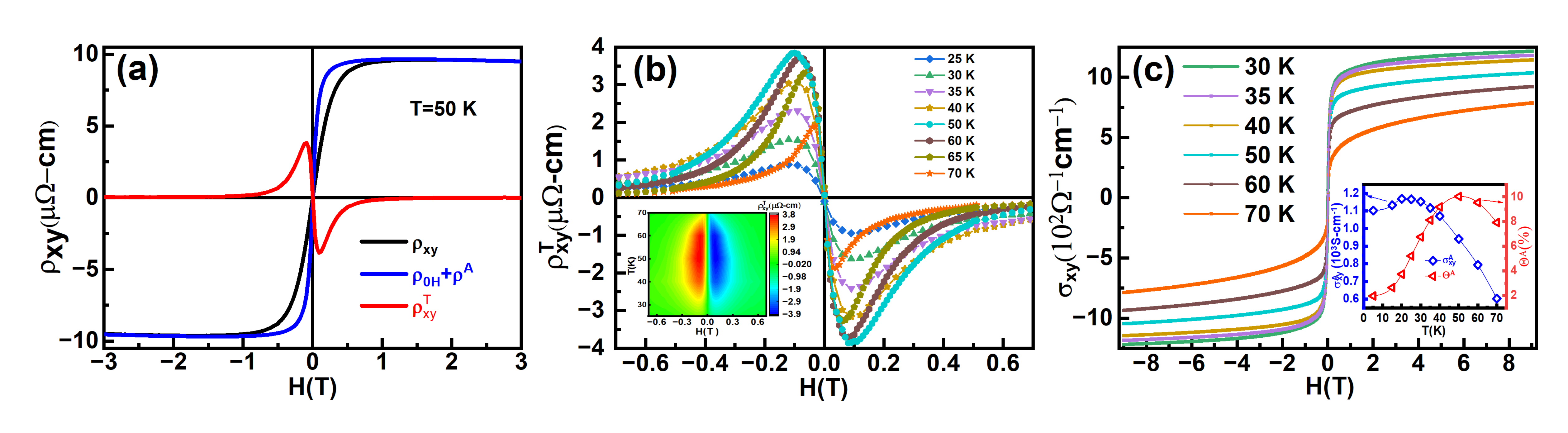}
\caption{(a) Variation of $\rho_{xy}$ (black curve), $\rho_{xy}^N$ + $\rho_{xy}^A$ (blue curve) and the topological hall resistivity, i.e.,  $\rho_{xy}^{T} = \rho_{xy}$ - ( $\rho_{xy}^N$ + $\rho_{xy}^A$) (red curve) with the applied field (H) at 50 K for Nd$_{4}$Sb$_{3}$, (b) Topological Hall resistivity ($\rho_{xy}^T$) vs. Field(H) at several temperatures, (Inset) contour plot of $\rho_{xy}^T$ on the H-T phase diagram, and (c) Hall conductivity ($\sigma_{xy}$)  vs. H;  (Inset) Variation of AHC (blue) and AHA (red)  as a function of T. } \label{TAHE}
\end{figure*}

The anomalous hall resistivity can be analyzed through its relation to the magnetization (M), expressed as $\rho_{xy}^A$ = R$_SM$, where $R_S =  S_H\rho_{xx}^{\beta}$ with $S_H$ being a field independent parameter. The value of $\beta$ enables us to identify different scattering mechanisms responsible for the observed AHE. If  $\beta = 1$, then $\rho_{xy}^A$ $\propto$ $\rho_{xx}$, which implies that the  dominant mechanism is the skew scattering. However, if $\beta = 2$, then $\rho_{xy}^A$ $\propto$ $\rho_{xx}^2$, which points to side jump or intrinsic mechanism as the dominant one responsible for AHE.\cite{id4} A linear fit to the Log$\rho_{xy}^A$ versus Log$\rho_{xx}$ data, as shown in Figure \ref{nmu-scal}(b),  yields a value of $\beta$ = 1.81, which indicates that the AHE mainly arises from the intrinsic scattering mechanism. The individual contributions of skew- and intrinsic-scattering mechanisms to the AHE can be estimated with the help of Tian-Ye-Jin (TYJ) scaling, which is of the form; $\rho_{xy}^A$ = $\alpha_{sk} \rho_{xx_0} + \sigma_{xy}^{int}\rho_{xx}^2$.\cite{shahi2024spin} Here, $\alpha_{sk}$ related to the skew scattering mechanism and $\sigma_{xy}^{int}$ corresponds to the intrinsic mechanism ($\rho_{xx_0}$ being the residual longitudinal resistivity). The value of $\alpha_{sk}$ = 0.024 obtained  from TYJ scaling analysis (Figure \ref{nmu-scal}(c)) is too small, indicating negligible contribution from skew scattering. On the other hand,  the value of $\sigma_{xy}^{int}$ turns out to be  990 S/cm, which is 93 $\%$ of total $\sigma_{xy}^A$ at T = 40 K implying that the dominant contribution comes from the intrinsic mechanism.  In general, it is very difficult to separate the side jump and Berry curvature parts. However,  an estimate of side-jump contribution, $\sigma_{xy,sj}^{int}$, can be obtained by employing the relation $\sigma_{xy,sj}^{int}$ = $e$/$ha$ (E$_{SOC}$/E$_F$), where $E_{SOC}$ is the spin orbit coupling energy.\cite{bhattacharya2025giant, bhattacharya2024spin, wang2018large}  Using the values parameters,  $E_{SOC}/E_F$ $\sim$ 0.01 for metallic systems \cite{wang2018large} and the lattice constant $a$ = 9.4053(4) $\text{\AA}$, the $\sigma_{xy,sj}^{int}$ is estimated to be 4.1 $S/{cm}$, which is quite small compared to 990 S/cm, the intrinsic part of AHE.  This clearly shows that AHE in the present case essentially arises from the Berry curvature. We performed the ab initio calculations for the electronic band structure described below to further confirms that the AHC contribution coming from the Berry curvature.

Figure \ref{TAHE} (a) displays field dependence of total symmetrized Hall resistivity ($\rho_{xy}$) at 50 K as black curve as well as sum of the OHE and AHE, i.e., $\rho_{xy}^{N}$ + $\rho_{xy}^{A}$ (= $R_0$H + $R_S$M ) as blue curve. It can be clearly seen from the figure that the sum of OHE and AHE alone cannot account for total $\rho_{xy}$. The difference between the two curves, i.e., $\rho_{xy}$ - ($\rho_{xy}^{N}$ + $\rho_{xy}^{A}$), shown as red curve, is the additional contribution to the Hall effect, which is the Topological Hall Effect (THE) denoted by $\rho_{xy}^{T}$. So the total hall resistivity can be expressed as \cite{lee2007hidden, kanazawa2011large}
\begin{equation}
\rho_{xy} = R_0H + R_SM + \rho_{xy}^T
\end{equation}
The topological Hall contribution usually vanishes at high fields where all the spins are fully polarized in the applied field direction. So, the THE term, $\rho_{xy}^T$, is zero or negligible at high fields due to the disappearance of the chiral spin texture. Therefore, one can write the total hall resistivity in the high field region as  $\rho_{yx}$/$H$ = R$_0$ + S$_H\rho_{xx}^2$M/H.  The linear fit to $\rho_{xy}$/H  vs $(S_H \rho_{xx}^2$M)/H data at high fields (see Figure S4 of the supporting information), gives the values of $S_H$ and $R_0$ as slope and intercepts. The $R_0$ value obtained this way is consistent with previously estimated values. The anomalous Hall coefficient, $S_H$, reflects the sensitivity of anomalous Hall current to magnetization. It is clearly seen that the temperature variation of $\sigma_{xy}^A$ and $M_s$ are qualitative similar,which is  reflected in the scale factor  $S_H$ .Here the $S_H$ obtained above remains nearly constant at 0.28 V$^{-1}$ over the entire range, confirming the insensitivity of AHE to impurity scattering.\cite{bhattacharya2025anomalous,kim2018large} Figure \ref{TAHE}(b) shows the variation of topological hall resistivity $\rho_{xy}^T$ as function of magnetic field for various temperatures. It shows negative peak for positive fields and positive peak  for negative fields. The maximum value is observed for 50 K and it decrease on either side of 50 K. Similar trends in THE was observed in other compounds as well and the maximum value of $\rho_{xy}^T$ $\sim$ -3.9 $\mu\Omega$-cm for Nd$_4$Sb$_3$  is much higher than those reported for Fe$_3$Sn$_2$\cite{li2019large},  Fe$_5$Sn$_3$\cite{li2020large}, Fe$_3$GeTe$_2$\cite{you2019angular} Mn$_3$Ga, \cite{id5} MnSi \cite{neubauer2009topological} and MnGe.\cite{kanazawa2011large} The H-T phase diagram constructed for topological Hall resistivity is shown in the inset of Figure \ref{TAHE}(b) which clearly brings out where THE appears in the temperature and field regimes.

Further, the anomalous hall angle (AHA), $\Theta^A$, which is defined as $\Theta^A$ = $\rho_{xy}^A$/$\rho_{xx}$, is another parameter that is used to characterize THE. The AHA specifies the amount of charge carriers that deviate from their applied current direction. The inset of Figure \ref{TAHE}(c) shows the variation of  $\Theta^A$ with T, which is similar to the variation of $\rho_{xy}^A$ shown in the inset of Figure \ref{nmu-scal}(c). The AHA also exhibits a peak with a value of 9.4\% around 60 K and drops off on either side of this temperature. It is inferred from several topological systems that the AHA value greater than 5$\%$ is a strong indication of non-trivial topological Berry curvature. For example, the AHA ($\Theta^A$) values for some of the topological materials are  Co$_3$Sn$_2$S$_2$ ($\Theta^A$ = 20$\%$),\cite{wang2018large} Fe$_3$GeTe$_2$ ($\Theta^A$ = 10$\%$),\cite{you2019angular}, Co$_2$MnAl ($\Theta^A$ = 20$\%$), \cite{li2020giant}, etc.

The Hall conductivities, both longitudinal ($\sigma_{xx}$) and Hall ($\sigma_{xy}$), can be calculated using the relations
\begin{equation}
 \sigma_{xy} = \frac{\rho_{yx}}{\rho_{yx}^2 + \rho_{xx}^2} \qquad \qquad \sigma_{xx}=\frac{\rho_{xx}}{\rho_{yx}^2 + \rho_{xx}^2}  \; .
\end{equation}
The variation of $\sigma_{xy}$ with magnetic field for different temperatures is shown in Figure \ref{TAHE}(c). The anomalous Hall conductivity (AHC), $\sigma_{xy}^A$, can be calculated as the intercept from a linear fit to high field data of $\sigma_{xy}$  vs H. The variation of $\sigma_{xy}^A$ with temperature is shown in the inset of Figure \ref{TAHE}(c). The maximum (AHC) was found to be 1170 $\Omega^{-1}$-$cm^{-1}$ at 30 K, which is comparable to the values found in other topological materials such as MgMn$_6$Ge$_6$\cite{du2026large}, GdMn$_6$Ge$_6$\cite{tao2025tunable} but much larger than the isostructural compound Gd$_4$Sb$_3$. \cite{id9} Magnetic Weyl semimetals, are generally anticipated to exhibit large anomalous Hall conductivity and anomalous Hall angle, arising from the Berry curvature induced by non trivial Band topology. The overall electrical transport results show that Nd$_4$Sb$_3$ exhibits large topological Hall response, intrinsic anomalous Hall resistivity and anomalous Hall Conductivity, all pointing to a strong contributions coming from topological Berry Curvature as well as a non-coplanar magnetic structure. These results are compared to the theoretical first principles electronic band structure calculations described below.

\subsection{{\it Ab-initio} Calculations}

\subsubsection{Electronic Structure and Spin Polarization}

Nd$_4$Sb$_3$ crystallizes in a non-centrosymmetric cubic structure with space group $I\bar{4}3d$, as illustrated in Figure~\ref{cstr}(a). As an initial reference for establishing its spin-polarized electronic structure, we considered the collinear ferromagnetic (FM) state with the magnetic moments oriented along different crystallographic directions, as shown in Figure~S5(a)--S5(c) of the supporting information. The corresponding relative energies are summarized in Table~S-I of the supporting information. Among the considered orientations, magnetization along the [001] direction yields the lowest total energy, identifying [001] as the preferred magnetization direction within the collinear FM state. We therefore employ this configuration as the reference state for analyzing the spin-resolved electronic structure and subsequently compare it with energetically competing canted magnetic configurations (presented in the next section).

The theoretically optimized lattice parameter of the primitive Nd$_4$Sb$_3$ cell is $a=b=c=4.69$~\AA. The electronic structure was calculated using the corresponding body-centered cubic unit cell, with the bulk band dispersion evaluated along the high-symmetry paths of the first Brillouin zone (BZ), as shown in Figure~\ref{cstr}(b). The calculated magnetic moment for the collinear FM state is $3.69~\mu_B$/Nd, which is in reasonable agreement with the previously simulated value of $3.27~\mu_B$/Nd \cite{legvold1980ep} and experimentally obtained value of $\sim3.58~\mu_B$/Nd. This consistency provides a useful validation of the underlying electronic-structure description before considering more complex magnetic configurations.

\begin{figure}[t]
\centering
\includegraphics[width=0.48\textwidth]{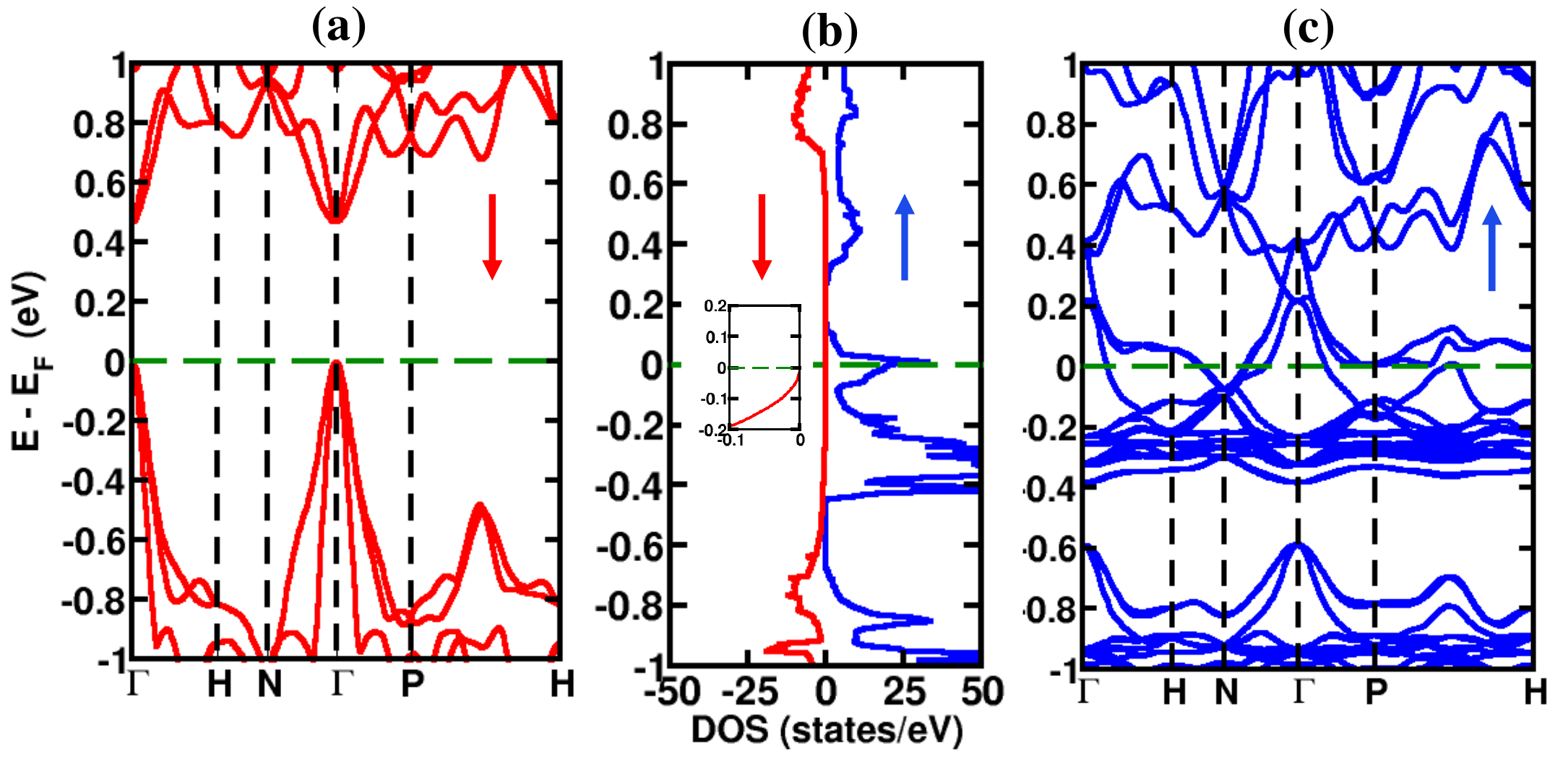}
\caption{Spin-resolved electronic band structure and density of states (DOS) of ferromagnetic (FM) Nd$_4$Sb$_3$ calculated within the GGA+$U$ approach. The inset shows an enlarged view of the DOS in the vicinity of the Fermi level (E$_F$), highlighting the pronounced spin asymmetry of the electronic states.}
\label{SRES}
\end{figure}

The spin-resolved electronic band structure and density of states (DOS), calculated within the GGA+$U$ approach, are presented in Figure~\ref{SRES}. A pronounced spin asymmetry is evident in the vicinity of the Fermi level. The spin-$\uparrow$ channel retains dispersive bands crossing E$_F$ and therefore exhibits a clear metallic character. In contrast, the spectral weight of the spin-$\downarrow$ channel is strongly suppressed around E$_F$, as also reflected in the spin-resolved DOS. In particular, the spin-$\downarrow$ bands exhibit a gap of approximately $0.47$~eV at the $\Gamma$ point, whereas the spin-$\uparrow$ states remain metallic. The resulting strong imbalance between the two spin channels establishes a highly spin-polarized electronic state in Nd$_4$Sb$_3$. This combination of a metallic majority-spin channel and a gapped minority-spin channel
indicates a strongly spin-polarized, half-metallic electronic
structure.

Such strong spin selectivity is important for the magnetotransport and topological properties discussed below. In a magnetic system with substantial spin polarization, changes in the underlying magnetic order can directly modify the spin-dependent band hybridization, spin-orbit-coupling-induced avoided crossings, and the associated Berry-curvature distribution near $E_F$. The collinear FM electronic structure therefore provides a physically transparent reference against which the consequences of magnetic canting can be assessed. This distinction is particularly relevant for Nd$_4$Sb$_3$, where the experimental results indicate that the intermediate-temperature ferromagnetic state evolves toward a more complex spin-canted magnetic state at low temperature \cite{ritter2010magnetic}. Consequently, a description restricted to the collinear FM state alone may not be sufficient to capture the electronic and topological characteristics of the low-temperature magnetic phase.
Motivated by this experimental evolution of the magnetic state, we therefore extend the first-principles analysis beyond the collinear FM approximation and explicitly examine several canted magnetic configurations. As discussed in the following subsection, this comparison reveals an energetically preferred canted state (Configuration-2), thereby establishing an intrinsic energetic tendency toward non-collinear magnetism and providing an important microscopic basis for examining the interplay between magnetic canting, real-space Berry-phase physics, and the momentum-space topology of Nd$_4$Sb$_3$.

\subsubsection{Magnetic Energetics and Canted Magnetic Ground State}

The experimental magnetic and magnetotransport measurements reveal a distinct evolution of the magnetic state of Nd$_4$Sb$_3$ upon cooling. Below the ferromagnetic transition at $T_{\mathrm{C}}\sim72$~K, the system develops a predominantly FM state, which undergoes a further spin reorientation around $T_{\mathrm{SR}}\sim42$~K. The concomitant reduction in magnetization, ZFC--FCW bifurcation, and pronounced changes in the magnetoresistance and Hall response indicate that the low-temperature magnetic state cannot be adequately described by a simple collinear FM configuration. This interpretation is also consistent with the previously reported spin-canted magnetic structure of Nd$_4$Sb$_3$ \cite{ritter2010magnetic}. Motivated by these experimental observations, we extended the first-principles calculations beyond the collinear FM approximation to examine whether the underlying magnetic energetics intrinsically favor a departure from collinearity.

For this purpose, several representative canted magnetic configurations were considered, as illustrated in Figure~S6(a)--S6(f) of the supporting information. While the collinear FM configuration contains Nd moments aligned along the $c$ axis, i.e., $(0,0,m_z)$, the canted configurations allow finite transverse components of the Nd moments. In particular, all three components $(m_x,m_y,m_z)$ of the Nd moments are finite in Configuration-2, allowing the magnetic structure to explore a genuinely three-dimensional spin configuration beyond the collinear FM manifold. The calculated relative energies and magnetic moments of the different configurations are summarized in Table~\ref{tab:config_energy}. Among the magnetic structures considered, Configuration-2 emerges as the lowest-energy state. Importantly, the collinear FM state lies only $\sim19$~meV/f.u. above Configuration-2, while Configuration-5 is separated by merely $\sim13$~meV/f.u. These relatively small energy differences reveal the presence of competing low-energy magnetic states in Nd$_4$Sb$_3$, rather than an overwhelmingly dominant magnetic configuration. Such a shallow magnetic-energy landscape is particularly relevant to the experimentally observed temperature-driven evolution of the magnetic structure. At intermediate temperatures below $T_{\mathrm{C}}$, the dominant ferromagnetic interactions can stabilize a predominantly FM state, whereas at lower temperatures the competition among exchange interactions, magnetic anisotropy, and antisymmetric interactions can favor a canted magnetic configuration. The calculated stabilization of Configuration-2 is therefore consistent with the experimentally observed spin reorientation at $T_{\mathrm{SR}}$ and provides a microscopic basis for the emergence of a more complex low-temperature magnetic state.

\begin{table}[t]
\centering
\caption{Calculated relative energies ($\Delta E$) and magnetic moments of the collinear FM and different canted magnetic configurations (shown in Figure S6 of supporting information) considered for Nd$_4$Sb$_3$. Configuration-2, which is found to be the lowest-energy state, is taken as the reference.}
\label{tab:config_energy}
\begin{tabular}{ccc}
\hline
Configurations & $\Delta E$ (meV/f.u.) & Magnetic moment ($\mu_B$/Nd) \\
\hline
FM & 19 & 3.69 \\
Configuration-1 & 46 & 2.44 \\
Configuration-2 & 0 & 3.27 \\
Configuration-3 & 234 & 3.12 \\
Configuration-4 & 54 & 3.23 \\
Configuration-5 & 13 & 3.25 \\
\hline
\end{tabular}
\end{table}

The calculated magnetic moment of Configuration-2 is $\sim3.27~\mu_B$/Nd, compared with $3.69~\mu_B$/Nd for the collinear FM state. The reduction in the net magnetic moment upon canting can naturally arise from the partial cancellation of differently oriented moment components and is qualitatively consistent with the experimentally observed modification of the magnetization across the spin-reorientation regime. More importantly, the energetic stabilization of Configuration-2 demonstrates that the non-collinear degrees of freedom are not merely imposed trial perturbations to the FM state, but constitute an energetically favorable solution within the first-principles description.
A particularly important consequence of Configuration-2 is its non-coplanar magnetic character. To quantify this feature, we evaluated the scalar spin chirality associated with different triplets of Nd moments,
\begin{equation}
\chi_{ijk}=\mathbf{S}_{i}\cdot
\left(\mathbf{S}_{j}\times\mathbf{S}_{k}\right),
\label{eq:scalar_chirality}
\end{equation}
where $\mathbf{S}_{i}$, $\mathbf{S}_{j}$, and $\mathbf{S}_{k}$ denote the magnetic moments on three Nd sites. In contrast to a collinear or coplanar magnetic configuration, for which $\chi_{ijk}$ vanishes, the calculated Configuration-2 exhibits finite local scalar spin chirality. For example, among the Nd spin triplets considered within the unit cell, values as large as $|\chi_{ijk}|\sim0.35~\mu_B^3$ are obtained. Equivalently, using normalized spin vectors, $\hat{\mathbf{S}}_i=\mathbf{S}_i/|\mathbf{S}_i|$, the corresponding dimensionless chirality,
\begin{equation}
\tilde{\chi}_{ijk}=
\hat{\mathbf{S}}_{i}\cdot
\left(\hat{\mathbf{S}}_{j}\times\hat{\mathbf{S}}_{k}\right),
\end{equation}
reaches $|\tilde{\chi}_{ijk}|\sim9.5\times10^{-3}$. The finite value directly demonstrates that the DFT-stabilized Configuration-2 is not simply a uniformly rotated ferromagnetic state but possesses local non-coplanar spin correlations.
The presence of finite scalar spin chirality provides an important microscopic connection between the calculated canted magnetic state and the experimentally observed topological Hall effect (THE). In the adiabatic picture, conduction electrons propagating through a non-coplanar magnetic background acquire an additional geometric phase determined by the solid angle subtended by the local magnetic moments. This real-space Berry phase can be represented as an effective emergent magnetic field acting on the itinerant carriers and can consequently generate an additional Hall response. Thus, unlike the collinear FM reference state, Configuration-2 contains a key microscopic ingredient required for a real-space topological contribution to the Hall effect: finite local scalar spin chirality arising from its non-coplanar magnetic arrangement.

This connection becomes particularly relevant when viewed together with the experimental results. The pronounced THE develops in the same low-temperature regime in which Nd$_4$Sb$_3$ undergoes spin reorientation and departs from the predominantly collinear FM state. Independently, the first-principles calculations identify Configuration-2 as the lowest-energy magnetic state and establish finite local scalar spin chirality in this configuration. The concurrence of these observations strongly suggests that the experimentally observed THE is closely connected with the development of non-collinear/non-coplanar magnetic correlations below the spin-reorientation transition. The calculated magnetic energetics and scalar spin chirality therefore provide a microscopic bridge between the experimentally observed magnetic reconstruction and the real-space Berry-phase physics relevant to the THE.

We emphasize, however, that the finite local scalar spin chirality obtained for Configuration-2 should not be interpreted as a direct microscopic simulation of the complete spatially extended topological spin texture inferred from the Hall measurements. The actual low-temperature magnetic state may involve longer magnetic length scales and spatial variations of the spin orientations that cannot be represented within the present two-formula-unit computational cell. Furthermore, the macroscopic emergent field depends not only on the magnitude of the local scalar chirality but also on the geometry, orientation, and collective arrangement of the relevant magnetic plaquettes. Establishing such an extended texture and its net emergent field would require large-scale atomistic-spin or micromagnetic simulations, which are beyond the scope of the present study. Accordingly, the present calculations do not constitute a direct theoretical evaluation of the measured topological Hall resistivity. Rather, they establish at the microscopic level that (i) a canted magnetic state is energetically favored over the simple collinear FM state and (ii) this state possesses finite local scalar spin chirality, thereby providing a microscopic basis for real-space Berry-phase physics in Nd$_4$Sb$_3$.

The theoretical picture emerging from these results therefore involves two complementary manifestations of topology. The first originates in \textit{real space}, where the energetically stabilized non-coplanar magnetic state carries finite local scalar spin chirality and provides a microscopic route toward an emergent Berry phase and the experimentally observed THE. The second originates in \textit{reciprocal space}, where the magnetic Weyl electronic structure generates pronounced momentum-space Berry curvature and consequently a large intrinsic anomalous Hall response. Importantly, these two aspects are intertwined through the underlying magnetic order: the transition from the collinear FM state to the canted Configuration-2 lowers the magnetic symmetry and reconstructs the electronic band topology, thereby modifying the number and distribution of Weyl crossings and the associated Berry-curvature response. Configuration-2 thus provides a natural microscopic link between the real-space and momentum-space topological phenomena observed in Nd$_4$Sb$_3$.

To identify the microscopic magnetic interactions responsible for this competing magnetic landscape and the stabilization of non-collinearity, we next analyze the isotropic and anisotropic exchange interactions together with the Dzyaloshinskii--Moriya interaction (DMI), followed by finite-temperature Monte Carlo simulations.

\begin{figure}[t]
\centering
\includegraphics[width=0.48\textwidth]{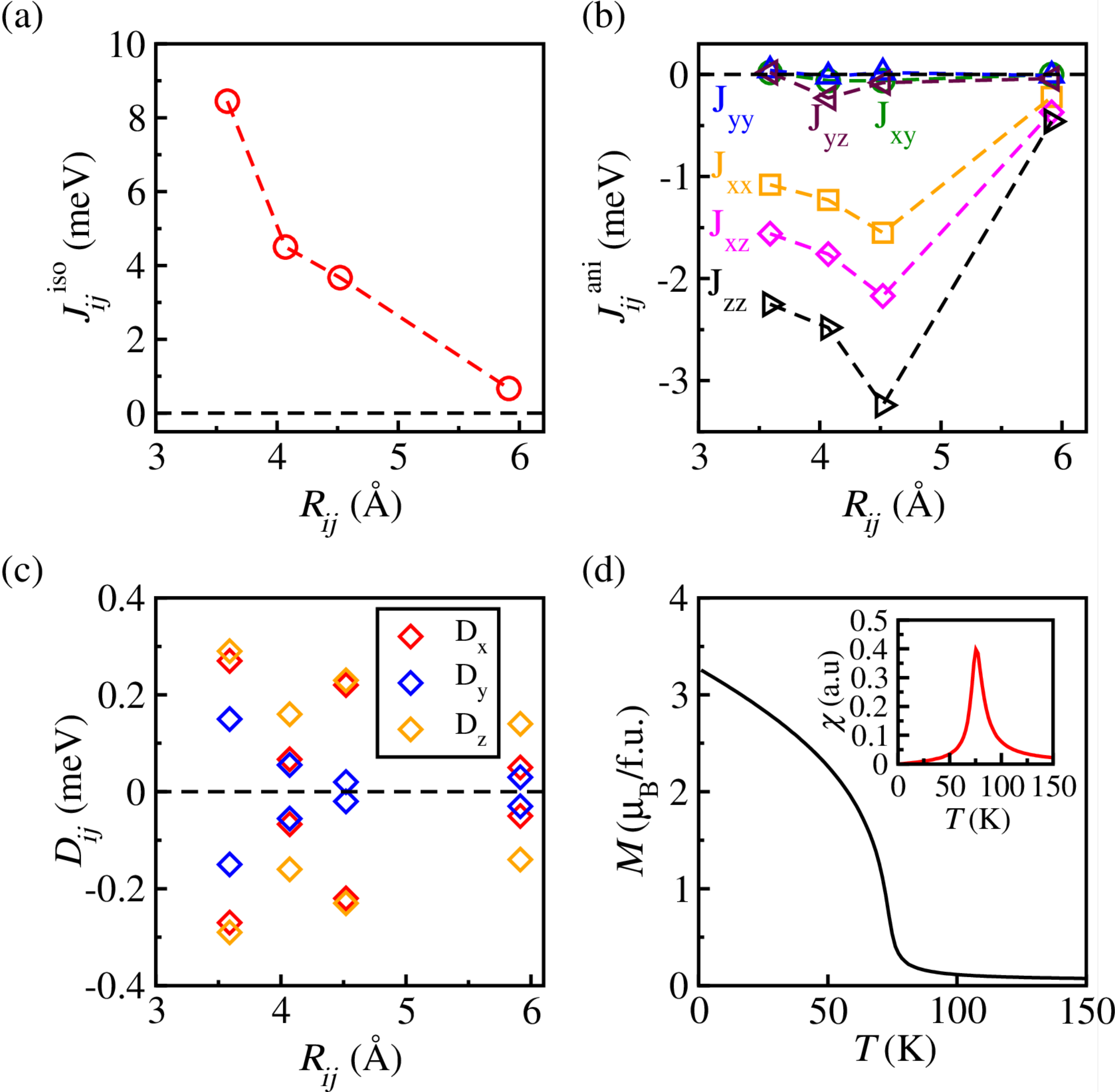}
\caption{Calculated magnetic interactions and finite-temperature magnetic response of Nd$_4$Sb$_3$ considering non-coplanar Configuration-2. (a) Isotropic exchange interactions ($J_{ij}^{\mathrm{iso}}$), (b) components of the symmetric anisotropic exchange tensor ($J_{ij}^{\mathrm{ani}}$), and (c) components of the Dzyaloshinskii--Moriya vector ($\mathbf{D}_{ij}$) as a function of the Nd--Nd interatomic distance ($R_{ij}$). (d) Monte Carlo simulated magnetization $M(T)$ obtained using the exchange parameters shown in (a-c). The inset shows the corresponding magnetic susceptibility $\chi(T)$. The characteristic magnetic transition occurs near $78$~K.}
\label{fig:T1}
\end{figure}

\subsubsection{Exchange Interactions and Finite-Temperature Magnetism}

To understand the microscopic interactions underlying the competing collinear and non-coplanar magnetic states of Nd$_4$Sb$_3$, we extracted the interatomic exchange parameters between the Nd moments from first-principles calculations using the Green's-function formalism implemented in the TB2J package \cite{he2021tb2j}. The resulting interactions were mapped onto the atomistic spin Hamiltonian given in Eq.~\ref{hamilton}, which contains the isotropic Heisenberg exchange $J_{ij}^{\mathrm{iso}}$, the symmetric anisotropic exchange tensor $J_{ij}^{\mathrm{ani}}$, and the antisymmetric Dzyaloshinskii--Moriya interaction (DMI), represented by the DM vector $\mathbf{D}_{ij}$. This decomposition is particularly relevant for Nd$_4$Sb$_3$, since the experimental results and magnetic energetics discussed above indicate a competition between a predominantly collinear FM state and a low-temperature canted/non-coplanar state.

The calculated isotropic exchange interactions [Figure~\ref{fig:T1}(a)] are predominantly ferromagnetic ($J_{ij}^{\mathrm{iso}}>0$) over the neighbor shells considered and decrease rapidly with increasing Nd--Nd separation. The dominant nearest-neighbor coupling is $J_1^{\mathrm{iso}}\sim8.4$~meV at $R_{ij}=3.59$~\AA, followed by interactions of approximately $4.5$, $3.7$, and $0.7$~meV at $R_{ij}=4.05$, $4.51$, and $5.91$~\AA, respectively. This hierarchy establishes strong short-range ferromagnetic correlations within the Nd sublattice and explains why a sizeable net FM component persists even in the canted low-temperature magnetic state. It is also consistent with the experimentally observed ferromagnetic ordering below $T_C$ [Figure ~2(a)].

The isotropic exchange alone, however, cannot account for the experimentally observed spin reorientation or the stabilization of the non-coplanar Configuration-2. The symmetric anisotropic exchange interactions [Figure ~\ref{fig:T1}(b)] provide an additional and appreciable magnetic-energy scale. Among the different tensor components, $J_{zz}$ and $J_{xz}$ are dominant, reaching values of approximately $-3.2$ and $-2.2$~meV, respectively. The $J_{xx}$ component is comparatively weaker, with a magnitude of approximately $1$--$1.6$~meV, whereas $J_{yy}$, $J_{xy}$, and $J_{yz}$ remain relatively small. The sizeable off-diagonal $J_{xz}$ component is particularly noteworthy, as it directly couples different Cartesian components of neighboring Nd moments and indicates that the principal axes of the exchange anisotropy are rotated with respect to the chosen crystallographic reference frame. Such anisotropic coupling provides a natural microscopic mechanism for competing moment orientations and facilitates departure from a perfectly collinear FM arrangement.
The non-centrosymmetric $I\bar{4}3d$ crystal structure further permits a finite Dzyaloshinskii--Moriya interaction. As shown in Figure ~\ref{fig:T1}(c), the calculated components of the DM vector, $D_x$, $D_y$, and $D_z$, are typically of the order of $0.10$--$0.30$~meV. Although the DMI is considerably weaker than the dominant isotropic exchange, it favors relative rotation of neighboring moments and therefore provides an additional microscopic source of canting. The resulting magnetic state can consequently be understood in terms of a hierarchy of competing interactions: strong short-range isotropic exchange establishes the predominantly ferromagnetic background, while sizeable anisotropic exchange and the weaker but finite DMI destabilize perfect collinearity and enable the development of a canted/non-coplanar magnetic state.

This hierarchy provides an important microscopic complement to the magnetic energetics and scalar-spin-chirality analysis discussed above. In particular, the dominant FM exchange explains why Configuration-2 retains a substantial net magnetization, whereas the orientation-dependent symmetric exchange and DMI provide microscopic interactions capable of generating the transverse moment components and non-coplanarity required for finite local scalar spin chirality. Thus, the same interaction hierarchy that accounts for the competing magnetic states also provides a microscopic route toward the real-space Berry-phase physics relevant to the experimentally observed topological Hall response.

To examine the finite-temperature consequences of these interactions, we performed classical Monte Carlo simulations using the complete exchange tensors $J_{ij}$ and DM vectors $\mathbf{D}_{ij}$ in the atomistic spin Hamiltonian of Eq.~\ref{hamilton}. For the exchange parameters evaluated using the non-coplanar Configuration-2 as the magnetic reference state, the calculated magnetization isotherm M(T) [Figure ~\ref{fig:T1}(d)] decreases sharply from its low-temperature value and approaches the disordered regime near $\sim78$~K. A corresponding anomaly is clearly visible in the magnetic susceptibility $\chi(T)$ [inset of Figure ~\ref{fig:T1}(d)], identifying a characteristic Monte Carlo transition temperature of approximately $78$~K. Remarkably, this temperature scale is close to the experimentally observed magnetic ordering temperature $T_C\sim72$~K, independently identified from the ZFC--FCW magnetization [Figure ~2(a)], Curie--Weiss analysis [inset of Figure ~2(a)], and the anomaly in $C_p(T)$ [Figure ~2(c)].

An instructive comparison is obtained by independently evaluating the exchange parameters around the collinear FM reference state and performing an otherwise analogous Monte Carlo simulation. In this case, the characteristic transition occurs at a higher temperature of approximately $95$~K (see Figure~S10 of the supporting information). The difference between the two calculated temperature scales is physically significant: allowing the underlying electronic and magnetic structure to adopt the non-coplanar Configuration-2, together with the corresponding anisotropic exchange and DMI, reduces the characteristic magnetic-ordering scale relative to that obtained from the idealized collinear FM reference state. This result demonstrates that the finite-temperature magnetic response of Nd$_4$Sb$_3$ is sensitive to the underlying magnetic configuration and cannot be described completely by an isotropic collinear-FM model.
The agreement between the $\sim78$~K Monte Carlo scale obtained for Configuration-2 and the experimental ordering temperature of $\sim72$~K is particularly encouraging, considering the approximations inherent in mapping the itinerant first-principles electronic structure onto a classical localized-spin Hamiltonian. More broadly, the comparison between the $\sim78$ and $\sim95$~K simulations emphasizes that the magnetic configuration itself significantly renormalizes the effective exchange landscape. This sensitivity is consistent with the experimentally observed magnetic reconstruction and with the relatively shallow energy landscape obtained from the first-principles total-energy calculations, where the collinear FM state lies only $\sim19$~meV/f.u. above Configuration-2.

Taken together, the exchange-interaction analysis, magnetic energetics, and finite-temperature simulations provide a coherent microscopic picture of magnetism in Nd$_4$Sb$_3$. Strong short-range ferromagnetic exchange establishes the dominant FM correlations, while sizeable anisotropic exchange and finite DMI introduce competing orientational tendencies that destabilize perfect collinearity and favor a non-coplanar magnetic state at low temperature. The resulting competition naturally supports the DFT-stabilized Configuration-2 and its finite local scalar spin chirality, thereby providing microscopic ingredients for the real-space topological Hall response. At the same time, the pronounced dependence of the calculated transition scale on the magnetic reference state highlights the intimate coupling between magnetic order and the underlying electronic structure. In the following subsection, we examine how this magnetic reconstruction modifies the reciprocal-space electronic topology, particularly the Weyl nodes and associated surface states.

\begin{figure}[t]
\centering
\includegraphics[width=0.48\textwidth]{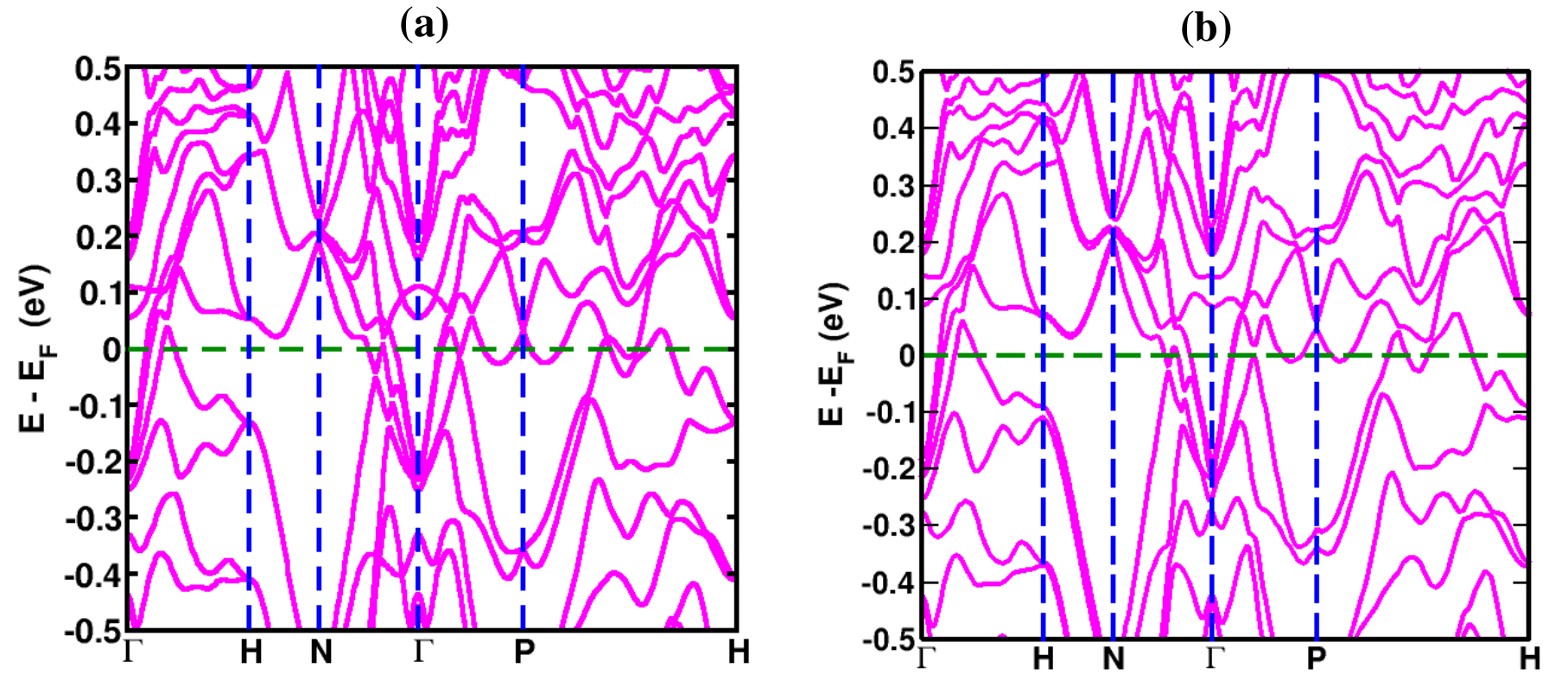}
\caption{Electronic band structure of (a) FM and (b) Configuration-2 of Nd$_4$Sb$_3$ calculated including the effect of spin-orbit coupling (SOC).}
\label{surf}
\end{figure}

\begin{figure*}[t]
\centering
\includegraphics[width=0.96\textwidth]{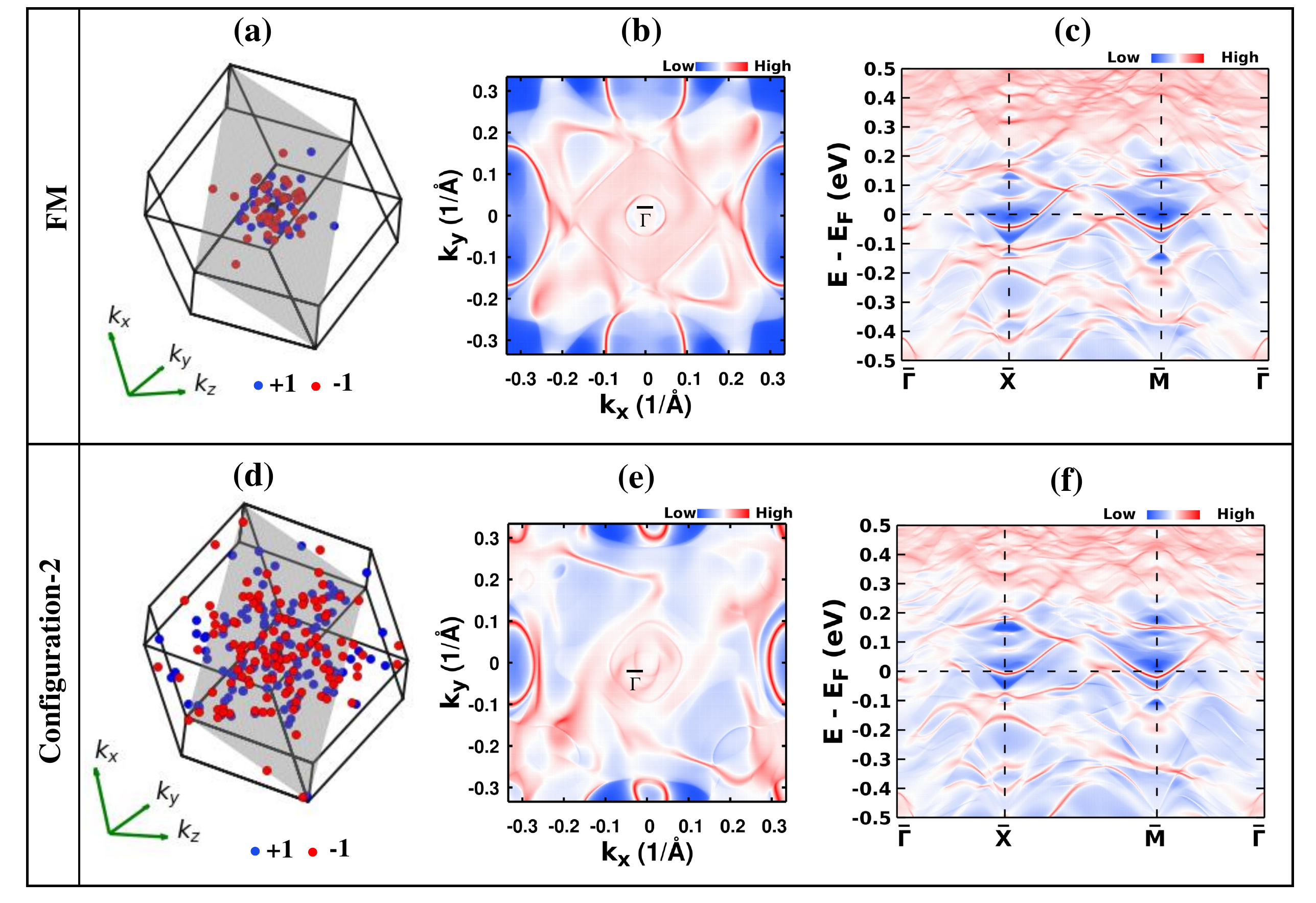}
\caption{For FM state of Nd$_4$Sb$_3$, (a) distribution of Weyl nodes within the bulk Brillouin zone in the energy range [-0.2 to 0.2] eV (b) Fermi surface, and (c) surface-state band dispersion along the [001] direction. (d,e,f) Same as (a,b,c) but for magnetic Configuration-2 of Nd$_4$Sb$_3$.}
\label{WND}
\end{figure*}

\subsubsection{Weyl Topology and Surface States}

Having established the presence of competing collinear and non-coplanar
magnetic states in Nd$_4$Sb$_3$, we next examine how the magnetic
configuration influences its reciprocal-space electronic topology.
Figure~\ref{surf} compares the bulk electronic structures of the
collinear FM state and the non-coplanar Configuration-2 in the presence
of spin--orbit coupling (SOC). For both magnetic configurations, a dense
set of dispersive bands and closely spaced crossings/avoided crossings
occurs in the vicinity of the Fermi level, particularly within the energy
window E-E$_F\sim[-0.2,0.2]$~eV. The orbital-projected density of states
(PDOS) for the FM state is presented in Figure~S7 of the supporting information,
showing that the electronic states around E$_F$ predominantly originate
from hybridized Nd-$d$ and Sb-$p$ states. A similar orbital character is
obtained for Configuration-2, indicating that the magnetic reconstruction
primarily modifies the spin-dependent hybridization and topology of the
low-energy bands rather than their dominant orbital character.

The Wannier-interpolated band structures  for both magnetic configurations, agreeing well with the DFT bands, are shown in
Figure~S8 of the supporting information, providing a reliable basis for the
topological analysis. A closer inspection of the SOC band structures
reveals numerous near-degeneracies and avoided crossings, particularly
along the N--$\Gamma$--P directions. SOC lifts several of these
degeneracies and opens relatively small gaps while preserving the
semimetallic character around E$_F$. Such SOC-induced avoided crossings
are particularly important for the transverse transport response, since
the Berry curvature can become strongly enhanced when occupied and
unoccupied bands approach each other closely in energy. Consistently, the
calculated Berry curvature along representative high-symmetry directions
(Figure~S9 of the supporting information) exhibits pronounced features in the same
energy range for both FM and Configuration-2.

Interestingly, the avoided crossings are generally more strongly
reconstructed in Configuration-2 than in the collinear FM state. In the
FM configuration, the collinear magnetic order retains a relatively
well-defined spin character of the electronic states. In contrast, the
finite transverse moment components in Configuration-2 mix different spin
components through SOC and modify the hybridization between nearby bands.
The non-coplanar magnetic order therefore acts not simply as a perturbation
to the FM electronic structure, but substantially reconstructs the
low-energy band-crossing landscape. This provides the first indication
that the experimentally observed evolution of the magnetic structure can
directly influence the reciprocal-space topology of Nd$_4$Sb$_3$.
To establish the topological character explicitly, we performed a
systematic search for isolated band crossings throughout the
3-dimensional (3D) Brillouin zone and evaluated their topological
chiralities. For the collinear FM state, we identify a total of
42 pairs of Weyl points (WPs) with opposite chirality. The
coordinates, energies relative to E$_F$, and chiralities of the identified
Weyl nodes are summarized in Tables~S-II and S-III of the supporting information. A
substantial fraction of these Weyl points occurs within
E-E$_F\sim[-0.2,0.2]$~eV, demonstrating that the Weyl topology is
energetically relevant to the experimentally probed electronic states.

More remarkably, the Weyl landscape changes substantially upon going from
the collinear FM state to the non-coplanar Configuration-2. For
Configuration-2, we identify 115 pairs of Weyl points with
opposite chirality, compared with only 42 pairs in the FM state. This
large increase demonstrates a pronounced sensitivity of the Weyl topology
to the underlying magnetic order. The reduction of magnetic symmetry and
spin mixing associated with the non-coplanar state lift some of the
symmetry relations present in the collinear FM configuration and permits
additional inequivalent band crossings to emerge throughout the Brillouin
zone. Thus, magnetic canting in Nd$_4$Sb$_3$ does considerably more than
modify the magnitude and orientation of the local magnetic moments: it
reconstructs the momentum-space topological landscape and substantially
increases the number of Weyl crossings.

Figure~\ref{WND}(a) shows the distribution of Weyl nodes within
E - E$_F\sim[-0.2,0.2]$~eV for the FM state. The Weyl points are distributed
over an extended region of momentum space, with a pronounced accumulation
around particular $k_z$ plane, producing a broken nodal-ring-like
distribution. \textcolor{black}{For Configuration-2, most of the Weyl points also remain concentrated in the $k_z$ plane, indicating that this plane hosts the majority of the band crossings as shown in Figure~\ref{WND}(d). However, the canting of the moments lead to the redistribution of some Weyl points into other planes of the Brillouin zone and resulting in a more dispersed distribution compared with the FM state.} Such a dense concentration of topological band crossings
near E$_F$ is particularly important because each Weyl point acts as a
source or sink of momentum-space Berry curvature. The collective presence
of a large number of Weyl nodes and nearby SOC-induced avoided crossings
can therefore generate an extended Berry-curvature landscape rather than
a response localized around only a few isolated points in momentum space.
As discussed in the following subsection, this provides a microscopic
basis for the unusually large intrinsic anomalous Hall response observed
experimentally.

The nontrivial bulk topology is further reflected in the surface electronic structure. The considered surface contains both
Nd- and Sb-terminated regions resulting from cleavage of the
Nd–-Sb bonds along the [001] direction. Figure~\ref{WND}(b) and \ref{WND}(c) show the
calculated $k_x$--$k_y$ Fermi surface and surface band dispersion,
respectively, for the (001) surface of the FM state in the presence of SOC.
The corresponding results for Configuration-2 are presented in
Figure~\ref{WND}(e) and \ref{WND}(f) respectively. \textcolor{black}{It is observed that the four-fold rotational symmetry of the bulk electronic structure is reduced to a two-fold rotational symmetry in the surface states, as shown in Figure~\ref{WND}(b) and \ref{WND}(e) for the FM and Configuration-2, respectively.}
Owing to the large number of
bulk Weyl nodes, several surface-derived states appear in the vicinity of
$\bar{\Gamma}$ and extend across the surface Brillouin zone. In particular,
the Fermi surface at E$_F$ [Figure~\ref{WND}(b)] exhibits extended
arc-like surface features, consistent with surface states associated with
the underlying Weyl-semimetallic bulk electronic structure.
The corresponding surface spectral dispersion along the
$\bar{\Gamma}$--$\bar{X}$--$\bar{M}$--$\bar{\Gamma}$ direction
[Figure~\ref{WND}(c)] contains several surface-state crossings in
the same energy range in which the bulk Weyl nodes are concentrated. A
prominent crossing occurs at approximately $0.09$~eV above E$_F$, while
another distinct crossing appears near $-0.02$~eV along the
$\bar{X}$--$\bar{M}$ direction. Additional surface-state features extend
throughout approximately $-0.2$ to $0.2$~eV. Configuration-2 likewise
exhibits surface-state crossings close to E$_F$, together with an
additional crossing around $\sim0.12$~eV [Figure~\ref{WND}(f)]. The simultaneous presence of bulk Weyl nodes and associated
surface-state features over this close energy window in the vicinity of E$_F$
provides complementary evidence for the Weyl-semimetallic character of
Nd$_4$Sb$_3$ in both magnetic configurations.

An important outcome of this analysis is therefore the robustness, yet
strong magnetic tunability, of the topological electronic structure. The
Weyl-semimetallic character survives in both the collinear FM and
non-coplanar magnetic states, whereas the detailed number and distribution
of Weyl nodes are dramatically reconstructed by magnetic canting. In
particular, the increase from 42 pairs of Weyl-points in the FM state to 115
pairs in Configuration-2 establishes a direct microscopic connection
between magnetic reconstruction and reciprocal-space topology. This result
is particularly significant when considered together with the finite
scalar spin chirality established for Configuration-2: the same
non-coplanar magnetic state that provides a microscopic route toward a
real-space Berry phase and the topological Hall effect also strongly
reorganizes the momentum-space Weyl topology.

Nd$_4$Sb$_3$ therefore provides an unusual setting in which real-space and
momentum-space topological phenomena are coupled through the magnetic
degree of freedom. The non-coplanar magnetic state supports finite local
scalar spin chirality relevant to the THE, while simultaneously generating
a dense Weyl landscape that acts as a source of momentum-space Berry
curvature. The latter connection can be tested quantitatively through the
intrinsic anomalous Hall conductivity, which we examine in the following
subsection.

\begin{figure*}[t]
\centering
\includegraphics[width=1.0\textwidth]{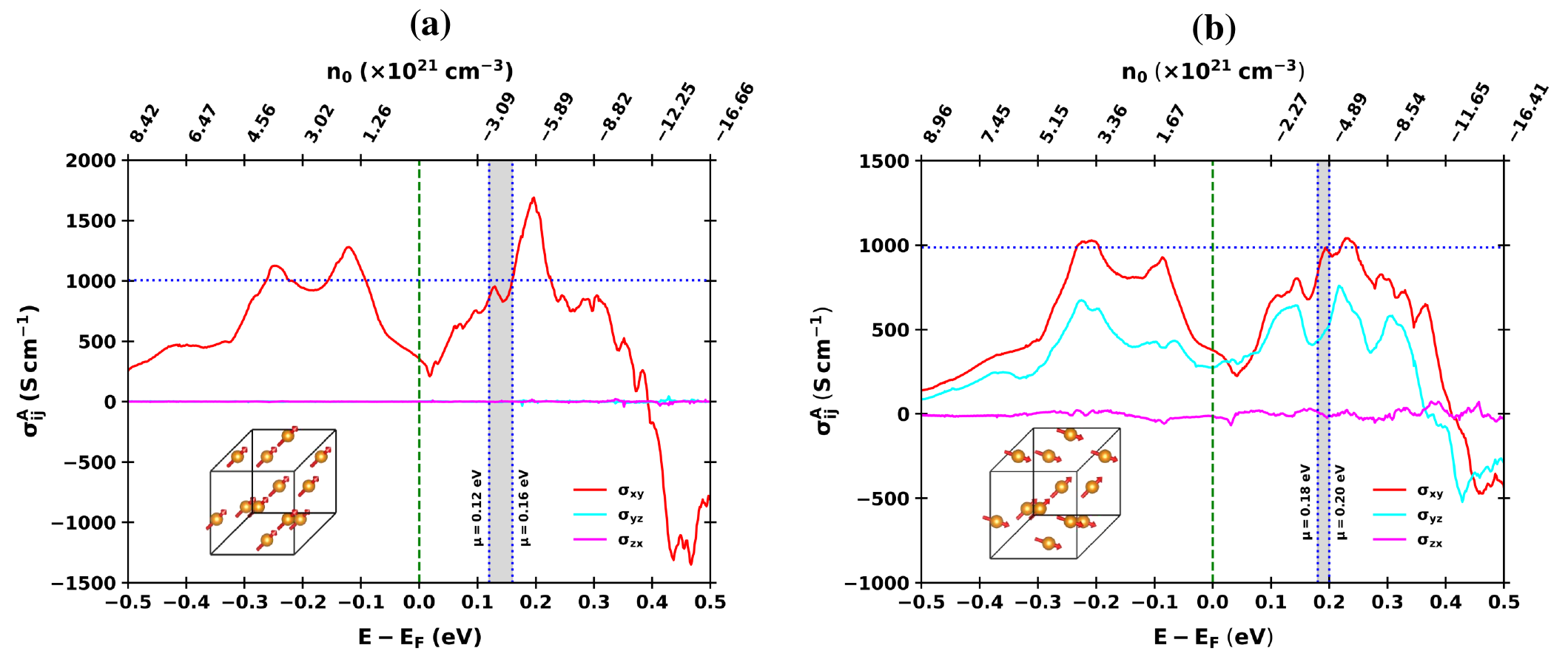}
\caption{The anomalous Hall conductivity ($\sigma_{ij}^A$) vs. chemical potential ($\mu$ = E - E$_{\text{F}}$) and the corresponding carrier concentration, n$_0$ (top horizontal axis) for (a) FM and (b) Configuration-2 magnetic states of Nd$_4$Sb$_3$. The inset shows the magnetic structure of FM and Configuration-2 states.}
\label{AHC-TH}
\end{figure*}

\subsubsection{Berry Curvature and Anomalous Hall Conductivity}

The dense distribution of Weyl points and SOC-induced avoided crossings
identified above provides a natural source of enhanced momentum-space
Berry curvature and, consequently, a large intrinsic anomalous Hall
conductivity (AHC). To quantify this response, we calculated the
band-resolved Berry curvature $\Omega^{z}_{n}(\mathbf{k})$ and the
intrinsic anomalous Hall conductivity $\sigma^{A}_{ij}$ using the
Berry-curvature formalism,\cite{wang2006ab,wang2018large}
\begin{equation}
\Omega^{z}_{n}(\mathbf{k}) =
-2\,\mathrm{Im}\sum_{m\neq n}
\frac{
\left\langle n\left|\frac{\partial H}{\partial k_a}\right|m\right\rangle
\left\langle m\left|\frac{\partial H}{\partial k_b}\right|n\right\rangle
}
{\left[E_m(\mathbf{k})-E_n(\mathbf{k})\right]^2},
\label{eq:berry}
\end{equation}
and
\begin{equation}
\sigma^{A}_{ij} =
-\frac{2\pi e^2}{h}
\int_{\mathrm{BZ}}
\frac{d^3k}{(2\pi)^3}
\sum_n f_n(\mathbf{k})\,\Omega^{z}_{n}(\mathbf{k}),
\label{eq:ahc}
\end{equation}
where $|n\rangle$ and $|m\rangle$ denote the eigenstates of the Hamiltonian
$H$ with corresponding eigenenergies $E_n$ and $E_m$, respectively,
and $f_n(\mathbf{k})$ is the Fermi--Dirac occupation factor. The inverse
quadratic dependence of the Berry curvature on the interband energy
separation makes the numerous Weyl crossings and SOC-induced
near-degeneracies identified above particularly important for the
intrinsic Hall response. Consistently, the calculated Berry curvature
along representative high-symmetry directions exhibits pronounced
features for both the FM and Configuration-2 states
(Figure~S9 of the supporting information).

The calculated energy-dependent AHC and corresponding carrier
concentrations for the FM and Configuration-2 states are shown in
Figure~\ref{AHC-TH}(a) and \ref{AHC-TH}(b), respectively, where
the chemical potential is defined as $\mu$ = E - E$_F$. A pronounced
dependence of the Hall response on both chemical potential and magnetic
configuration is immediately evident. For the collinear FM state, the
AHC is dominated by the $\sigma^{A}_{xy}$ component, consistent with
the preferred magnetization along the $z$-direction. At the nominal
DFT Fermi level ($\mu=0$), the calculated
$\sigma^{A}_{xy}$ is approximately $357$~S~cm$^{-1}$.
The Hall response becomes more complex in the non-coplanar
Configuration-2. In addition to a comparable
$\sigma^{A}_{xy}\sim377$~S~cm$^{-1}$ at $\mu=0$, a sizeable
$\sigma^{A}_{yz}\sim275$~S~cm$^{-1}$ is obtained. The appearance of
multiple sizeable off-diagonal conductivity components reflects the
lower magnetic symmetry of the non-coplanar state. Magnetic canting
mixes the spin components of the electronic states and reconstructs
the Weyl-node and avoided-crossing landscape, thereby redistributing
the momentum-space Berry curvature among different crystallographic
directions. Thus, the change from the collinear FM state to
Configuration-2 affects not only the magnitude but also the tensorial
character of the intrinsic anomalous Hall response.

Although the calculated $\sigma^{A}_{xy}$ values at the nominal DFT
Fermi level are smaller than the experimentally observed intrinsic
AHC, the Hall conductivity exhibits a remarkably strong dependence
on chemical potential. For the FM state,
$\sigma^{A}_{xy}$ increases rapidly within
$\mu\sim120$--$160$~meV, reaching approximately
$880$--$1006$~S~cm$^{-1}$. At $\mu=0.16$~eV,
$\sigma^{A}_{xy}$ reaches $\sim1006$~S~cm$^{-1}$, remarkably close
to the experimentally extracted intrinsic value of
$\sim990$~S~cm$^{-1}$. For Configuration-2, the region of enhanced
AHC is shifted slightly toward higher chemical potential,
$\mu\sim180$--$200$~meV. In this range,
$\sigma^{A}_{xy}$ reaches approximately $987$~S~cm$^{-1}$, while
$\sigma^{A}_{yz}$ simultaneously increases to approximately
$497$~S~cm$^{-1}$ at $\mu=0.195$~eV. The rapid enhancement of the
AHC over a relatively narrow energy window demonstrates the highly
energy-selective nature of the Berry-curvature distribution generated
by the Weyl crossings and SOC-induced band reconstruction.

A particularly important comparison with experiment is obtained by
considering the carrier concentration simultaneously with the AHC.
At the nominal Fermi level, the calculated intrinsic carrier
concentration for the FM state is
$n_0=2.54\times10^{20}$~cm$^{-3}$. Upon shifting the chemical
potential into the range $\mu\sim120$--$160$~meV, the calculated
electron concentration increases to
$|n_e|\sim(3.74$--$4.75)\times10^{21}$~cm$^{-3}$, which encompasses
the experimentally determined value of
$|n_e|\sim4.65\times10^{21}$~cm$^{-3}$. Importantly, this is the
same carrier-density/chemical-potential regime in which the calculated
intrinsic AHC becomes comparable to experiment.
At the experimentally measured carrier concentration,
$|n_e|\sim4.65\times10^{21}$~cm$^{-3}$, the calculated
$\sigma^{A}_{xy}$ reaches approximately $968$~S~cm$^{-1}$ for the
FM state and $\sim983$~S~cm$^{-1}$ for Configuration-2. Both values
are in close agreement with the experimentally extracted intrinsic
AHC of $\sim990$~S~cm$^{-1}$, with Configuration-2 providing the
closest correspondence. This agreement is particularly significant
because it does not arise simply from selecting a chemical potential
at which the calculated Hall conductivity matches experiment. Rather,
the experimentally measured carrier density independently identifies
the relevant electronic filling, and at this filling the calculated
Berry-curvature-driven AHC quantitatively reproduces the experimental
intrinsic Hall response. The simultaneous agreement of these two
independent quantities provides strong evidence that the large AHE in
Nd$_4$Sb$_3$ is predominantly governed by the intrinsic
momentum-space Berry curvature of its magnetic Weyl electronic structure. \textcolor{black}{In addition, the pronounced variation of $\sigma^{A}_{ij}$ with chemical potential highlights the possibility
of controlling the anomalous Hall response through carrier tuning. The upward shift of the chemical potential further indicates that electron-type carriers dominate in the relevant energy range of Nd$_4$Sb$_3$.}

The comparison between the two magnetic configurations provides an
additional insight. Both FM and Configuration-2 retain a large
intrinsic AHC at the experimentally relevant carrier density,
demonstrating that the strong Berry-curvature response is a robust
property of the magnetic Weyl electronic structure. Nevertheless,
Configuration-2 yields an AHC somewhat closer to the experimental
intrinsic value and additionally produces sizeable Hall-conductivity
components that are strongly suppressed in the collinear FM state.
Together with the substantial increase in the number of Weyl-point
pairs from 42 in the FM state to 115 in Configuration-2, these results
demonstrate that magnetic non-coplanarity strongly reconstructs the
momentum-space Berry-curvature landscape.

It is important to distinguish this intrinsic anomalous Hall response
from the topological Hall effect discussed experimentally. The
$\sigma^{A}_{ij}$ calculated here originates from the momentum-space
Berry curvature of the electronic bands and therefore provides a
microscopic description of the intrinsic AHE. In contrast, the
additional topological Hall contribution is associated with a
real-space Berry phase generated by non-coplanar magnetic correlations.
As demonstrated above, the energetically preferred Configuration-2
possesses finite local scalar spin chirality, thereby providing a
microscopic route toward this real-space contribution. The present
results therefore reveal two complementary Berry-phase mechanisms in
Nd$_4$Sb$_3$: momentum-space Berry curvature associated with the
magnetic Weyl electronic structure accounts quantitatively for the
large intrinsic AHC, whereas the non-coplanar magnetic state and its
finite scalar spin chirality provide the microscopic ingredient
required for the experimentally observed THE.
In other words, magnetic
order serves as the common control parameter connecting real- and
reciprocal-space topology: the low-energy non-coplanar state generates
finite scalar spin chirality in real space while simultaneously
reconstructing the Weyl-node distribution and momentum-space Berry
curvature. This coexistence provides a natural microscopic basis for
the simultaneous observation of a pronounced topological Hall effect
and a large intrinsic anomalous Hall conductivity in Nd$_4$Sb$_3$.

\section{Conclusion}

In summary, we establish Nd$_4$Sb$_3$ as a highly spin-polarized magnetic Weyl semimetal exhibiting intertwined anomalous and topological Hall responses. Nd$_4$Sb$_3$ orders ferromagnetically below $T_{\mathrm{C}}\sim72$~K and undergoes a spin-reorientation transition near $T_{\mathrm{SR}}\sim42$~K. The latter is accompanied by a pronounced topological Hall response reaching $\rho_{xy}^{\mathrm{T}}\sim3.9~\mu\Omega$\,cm, while the anomalous Hall conductivity reaches $\sim1170$~S\,cm$^{-1}$, with an experimentally extracted intrinsic contribution of $\sim990$~S\,cm$^{-1}$.

First-principles calculations identify an energetically preferred non-coplanar magnetic state with finite local scalar spin chirality, providing a microscopic route toward the real-space Berry phase underlying the topological Hall effect. The calculated exchange interactions reveal that strong ferromagnetic exchange, together with anisotropic exchange and finite Dzyaloshinskii--Moriya interaction, provides the microscopic ingredients for competing collinear and canted magnetic states. The magnetic reconstruction also profoundly modifies the reciprocal-space topology: the number of Weyl-point pairs increases from 42 in the collinear FM state to 115 in the non-coplanar state. At the experimentally determined carrier concentration, the calculated intrinsic AHC reaches $\sim983$~S\,cm$^{-1}$ for the non-coplanar state, in close agreement with the experimental intrinsic value of $\sim990$~S\,cm$^{-1}$.

Our results therefore reveal two distinct but magnetically intertwined Berry-phase mechanisms in Nd$_4$Sb$_3$: real-space scalar spin chirality provides the microscopic basis for the topological Hall effect, whereas momentum-space Berry curvature associated with the reconstructed Weyl electronic structure accounts for the large intrinsic anomalous Hall response. This coupling between magnetic reconstruction, spin chirality, and Weyl topology establishes Nd$_4$Sb$_3$ as a promising platform for magnetically tunable topological transport and Berry-phase-driven functionalities.

\section{Acknowledgment}
Subhasmita Kar acknowledges the Praganak (HPC)
Supercomputing facility at IIT Bombay for providing the computational resources used in this work. D.D. acknowledges PARAM Rudra, IIT Bombay for providing HPC facilitiy under the National Supercomputing Mission.


\bibliography{Ref}
\newpage
\begin{widetext}

\begin{center} {\Huge{\bf Supporting Information}} \end{center}
\setcounter{table}{0}
\renewcommand{\thetable}{S-\Roman{table}}

\vskip 0.3in

Here, we present details of Rietveld refinement of room temperature XRD (Fig.\ref{fig:XRD}), real and imaginary part of susceptibility data (Fig.\ref{fig:acx}), magnetic part of entropy and specific heat (Fig.\ref{fig:SM}), $\rho_{xy}$/H  vs. $(\rho_{xx}^2$M)/H plots(Fig.\ref{fig:THF}), various ferromagnetic configurations with different magnetization axes (Fig.\ref{fig:S5}), various non-collinear/non-coplanar magnetic configurations and their corresponding relative energies, projected density of states for FM Nd$_4$Sb$_3$ (Fig.\ref{fig:canted}), DFT and Wannierized band structures (Fig.\ref{fig:pdos}), Berry curvature (Fig.\ref{fig:band1}), Fermi surface and surface band dispersion plots (Fig.\ref{fig:Berry1}), Cartesian co-ordinates, energy position and the chirality of the Weyl points of the  FM phase, and calculated magnetic interactions and the Montecarlo simulated Magnetization (M) and susceptibility ($\chi$)  vs. T plots for FM Nd$_4$Sb$_3$ (Fig.\ref{fig:col}).

\setcounter{figure}{0}
\renewcommand{\thefigure}{S\arabic{figure}}

\begin{figure*}[h!]
    \centering
    \includegraphics[width=0.7\textwidth]{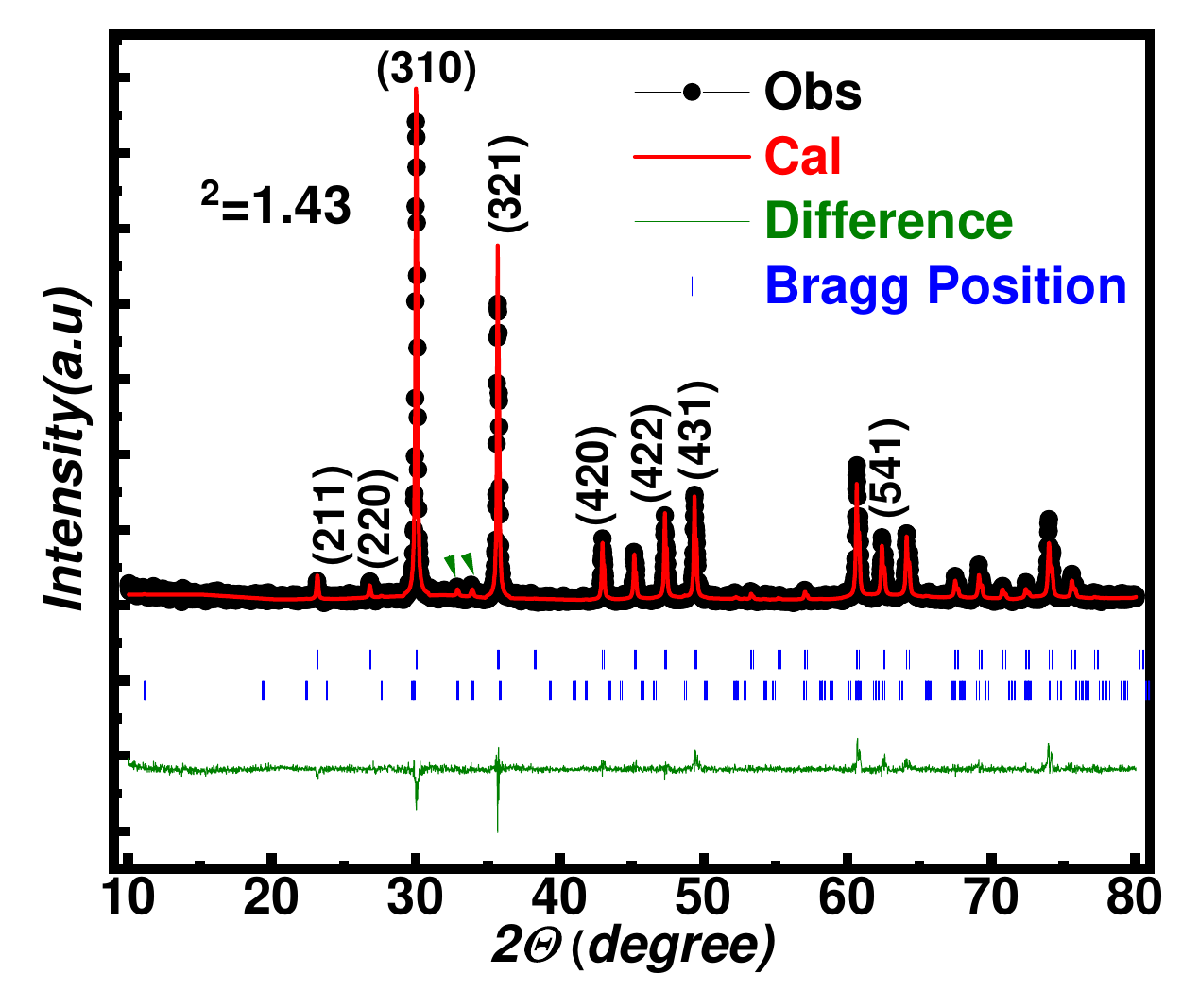}
    \caption {Room temperature (RT) powder XRD pattern and Rietveld refinement for Nd$_4$Sb$_3$. Small green arrows denote a minor impurity phase of Nd$_5$Sb$_3$, which is identified to be about 2.2$\%$ from the refinement}
    \label{fig:XRD}
\end{figure*}
\begin{figure*}[h!]
    \centering
\begin{minipage}[b]{0.48\textwidth}
    \centering
    \includegraphics[width=0.7\linewidth]{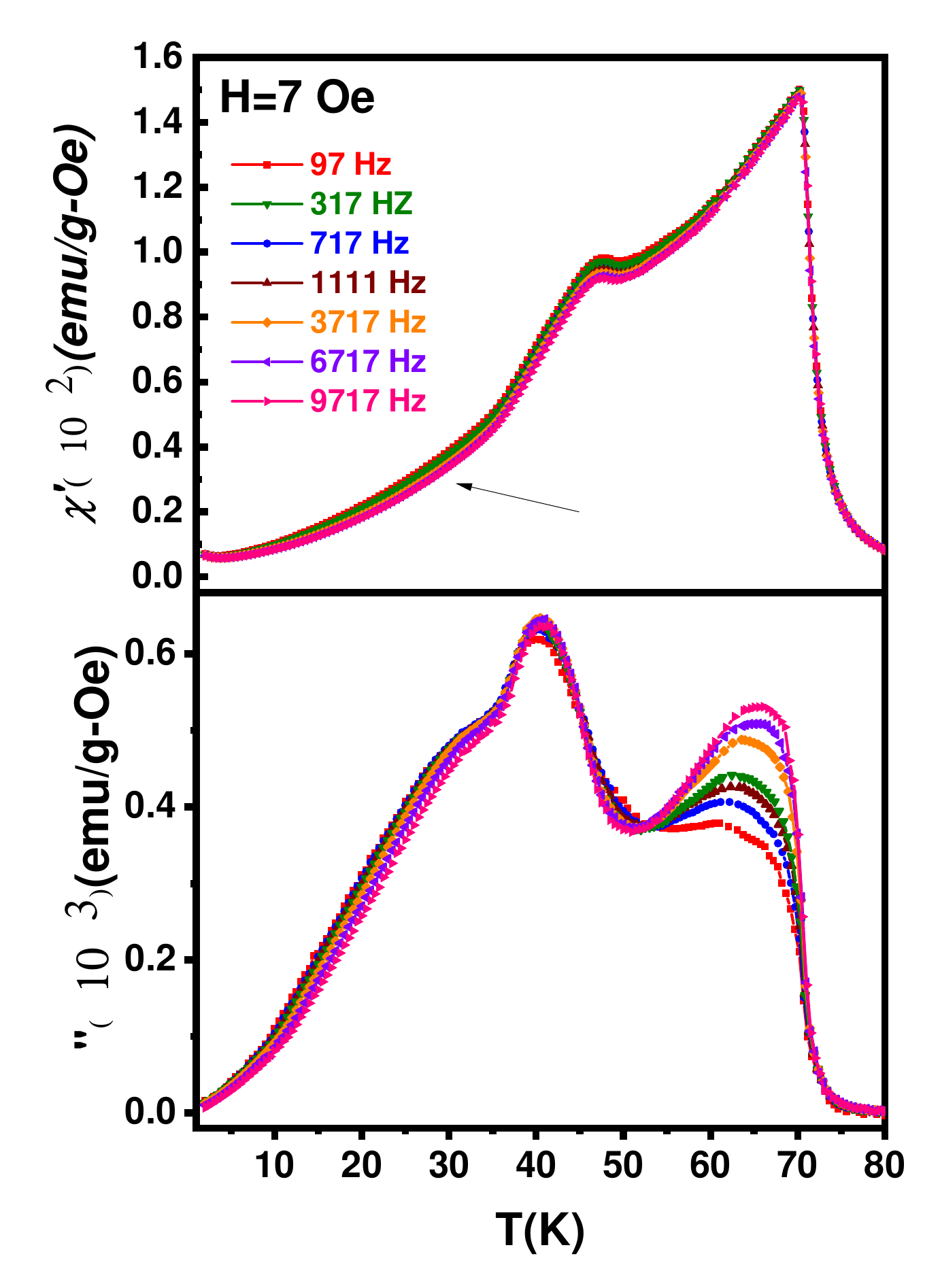}
    \caption{Real ($\chi^{\prime}$) and imaginary ($\chi^{\prime\prime}$) parts of ac-susceptibility measured as function of temperature  at several frequencies.} \label{fig:acx}
\end{minipage}
\hspace{0.5cm}
\begin{minipage}[b]{0.48\textwidth}
    \centering
    \includegraphics[width=\linewidth]{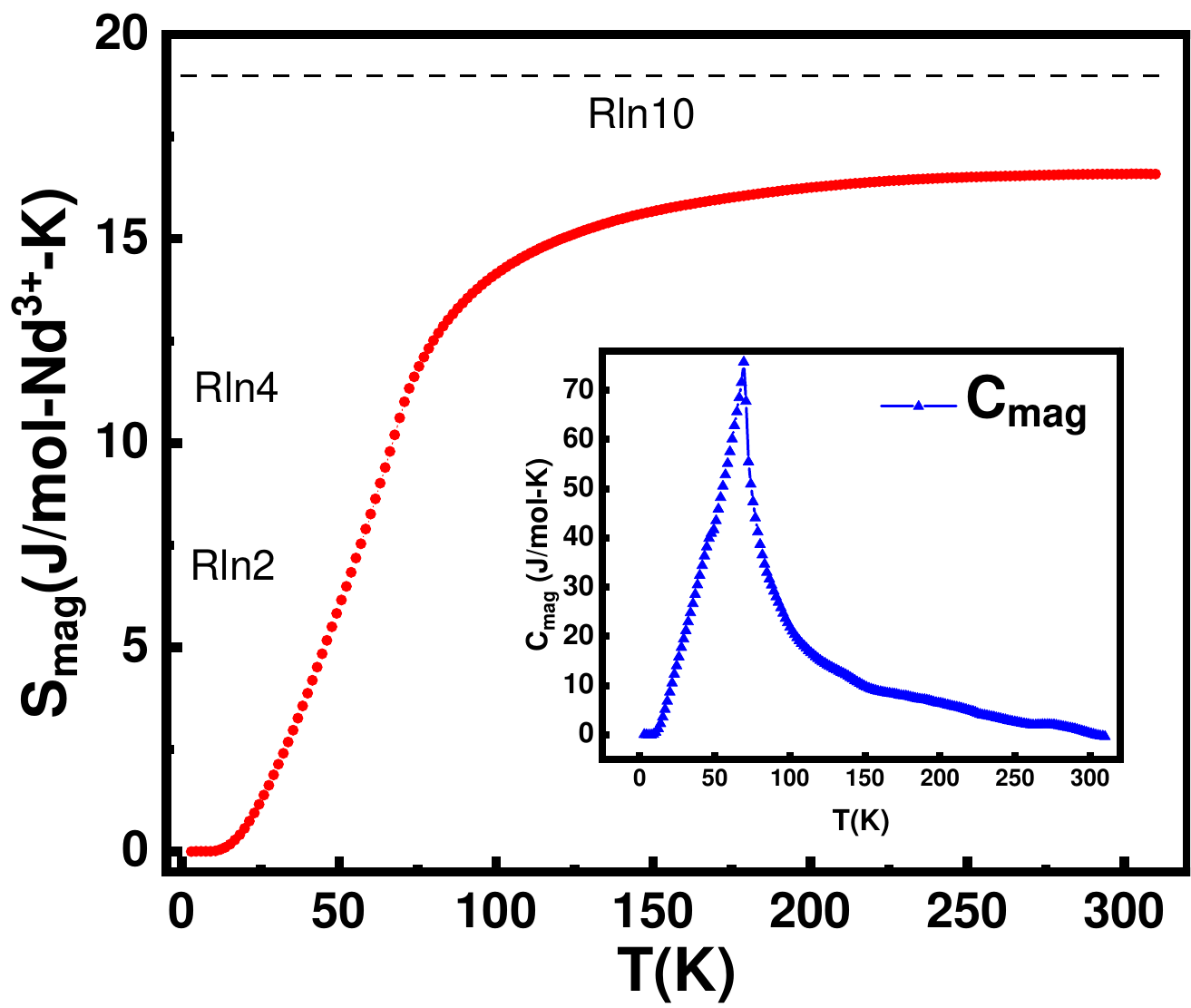}
    \caption{Magnetic part of entropy (S$_{\text{mag}}$) vs. T; (Inset) Magnetic part of specific heat (C$_{\text{mag}}$) vs. T.}
    \label{fig:SM}
\end{minipage}
\end{figure*}

\begin{figure*}[h]
    \centering
    \includegraphics[width=0.7\textwidth]{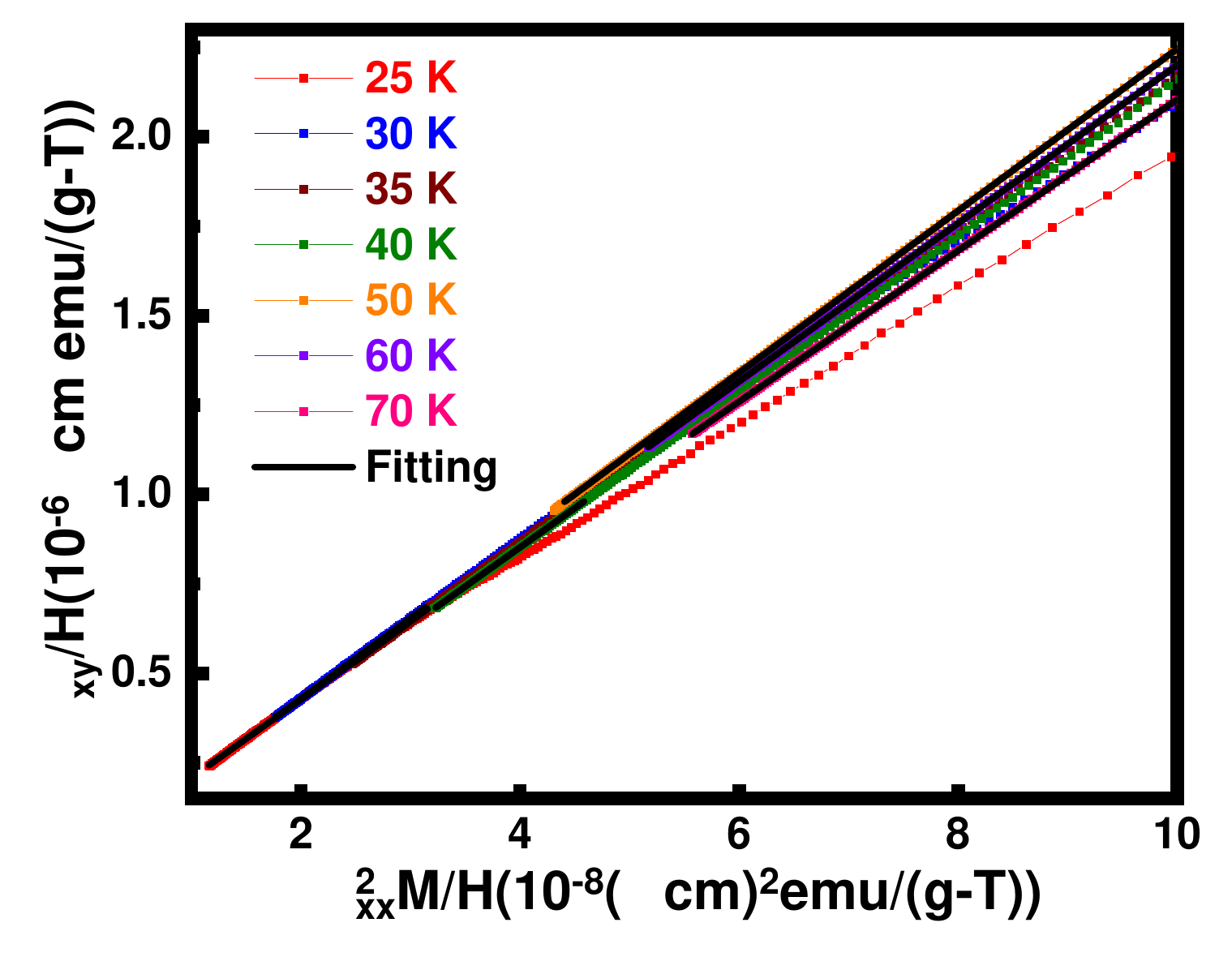}
    \caption {Plot of $\rho_{xy}$/H  vs. $(\rho_{xx}^2$M)/H  for all $\rho_{xy}$ isotherms over the entire measured temperature range between 25 K to 70~K}
    \label{fig:THF}
\end{figure*}

\begin{figure*}[t]
    \centering
    \includegraphics[width=0.8\textwidth]{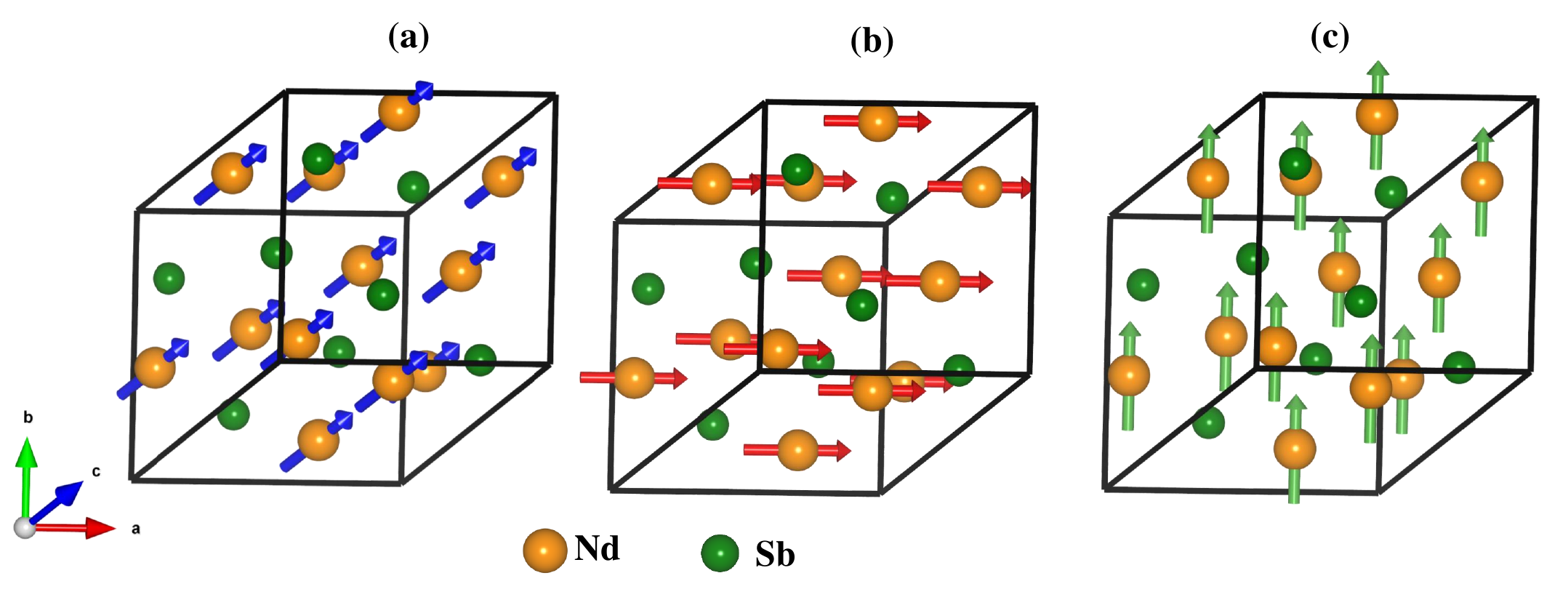}
    \caption {Ferromagnetic configurations of Nd$_4$Sb$_3$ with the Nd-spins oriented along (a) [001], (b) [100], and (c) [010] directions.}
    \label{fig:S5}
\end{figure*}
\vspace{1cm}

\begin{table}[h]
\centering
\caption{Relative energy ($\Delta$E) of the FM configuration with respect to different crystallographic directions [uvw]. [001] is the lowest energy direction and is taken as the reference state.}
\label{tab:CE}
\begin{tabular}{cc}
\hline
Magnetization Axis   \hskip 3cm &   \hskip 3cm $\Delta$E  \\

[uvw]  \hskip 3cm & \hskip 3cm (meV/f.u.) \\ \hline

[001]  \hskip 3cm &    \hskip 3cm 0.0  \\

[010]   \hskip 3cm &   \hskip 3cm 3.3  \\

[100]    \hskip 3cm &   \hskip 3cm 4.9  \\ \hline
\end{tabular}
\end{table}

\begin{justify} To understand the magnetic ground state in Nd$_4$Sb$_3$, we have considered three ferromagnetic (FM) configurations along [001], [100], and [010] directions, as depicted in Fig.~\ref{fig:S5}(a)-(c). Among the considered configurations, the orientation of the moments along the [001] direction exhibits the lowest energy, indicating that it is the preferred easy-magnetization axis for Nd$_4$Sb$_3$.
\end{justify}

\begin{figure*}[h]
    \centering
    \includegraphics[width=0.8\textwidth]{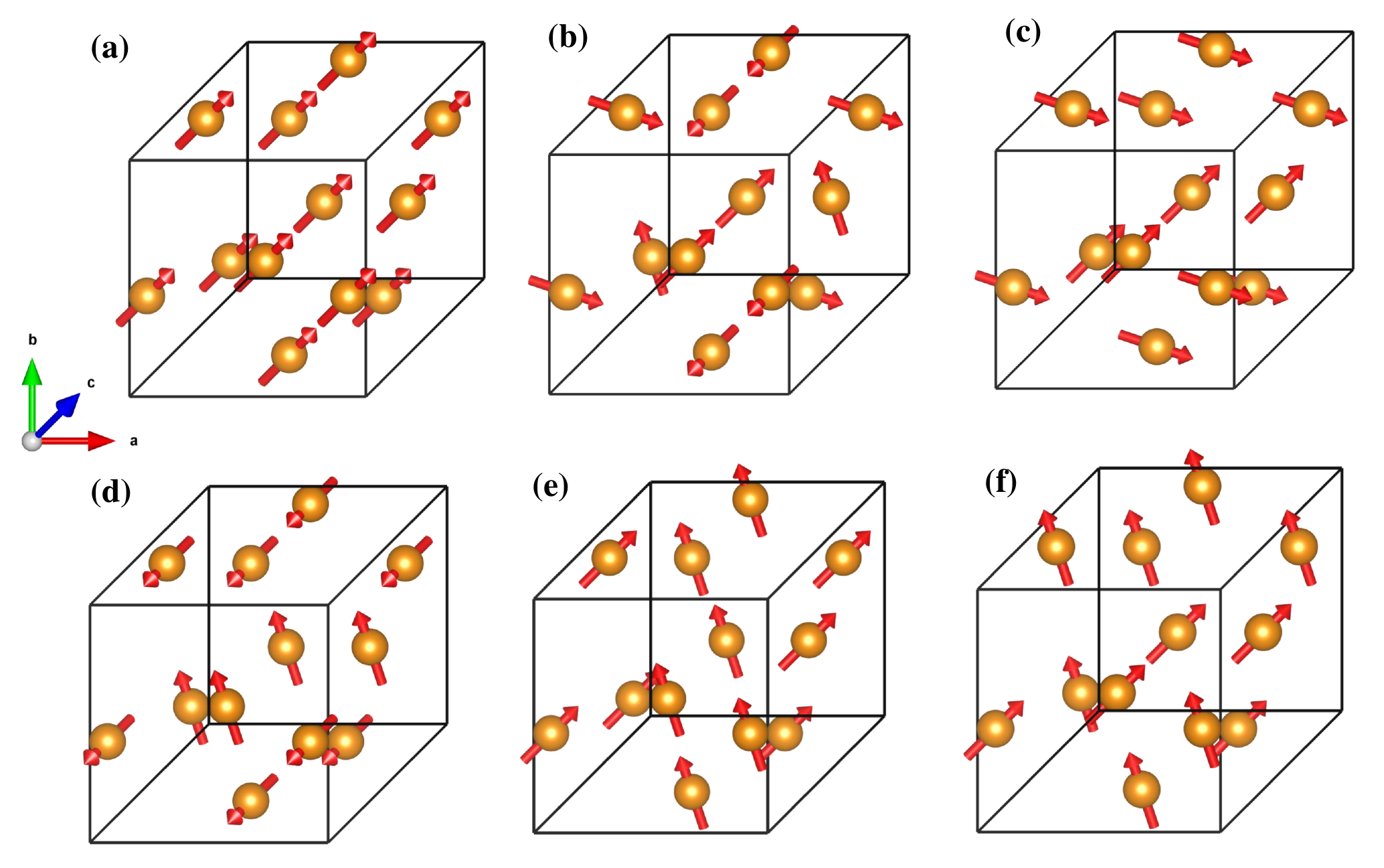}
    \caption {Various magnetic configurations of Nd$_4$Sb$_3$ (a) FM, and non-collinear/coplanar (b) Configuration-1, (c) Configuration-2, (d) Configuration-3, (e) Configuration-4, and (f) Configuration-5. Only Nd-atoms are displayed for clarity. }
    \label{fig:canted}
\end{figure*}

\begin{figure*}[h]
    \centering
    \includegraphics[width=0.8\textwidth]{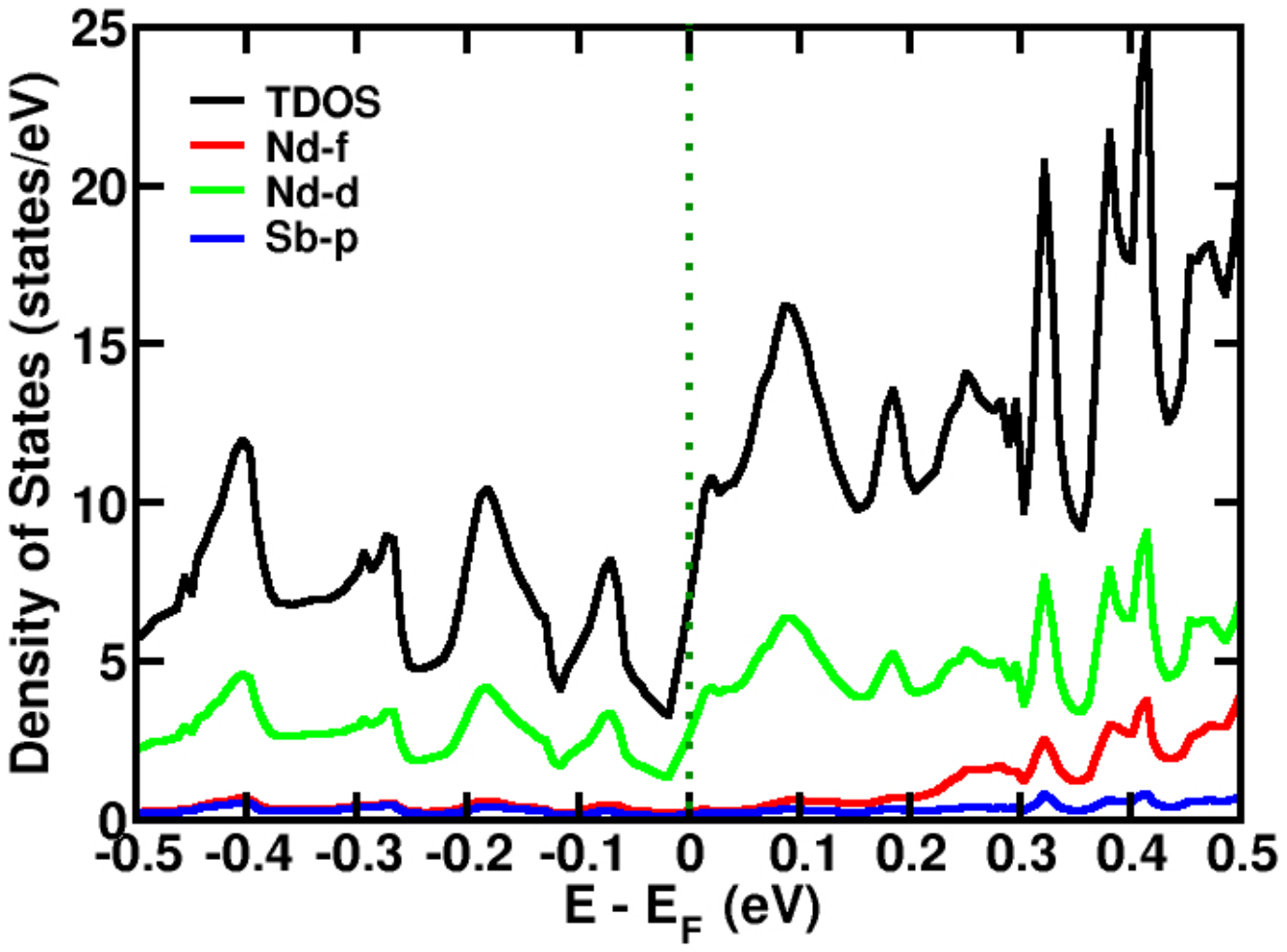}
    \caption {Electronic projected density of states (PDOS) for FM Nd$_4$Sb$_3$ without SOC.}
    \label{fig:pdos}
\end{figure*}

\begin{figure*}[h]
    \centering
    \includegraphics[width=0.8\textwidth]{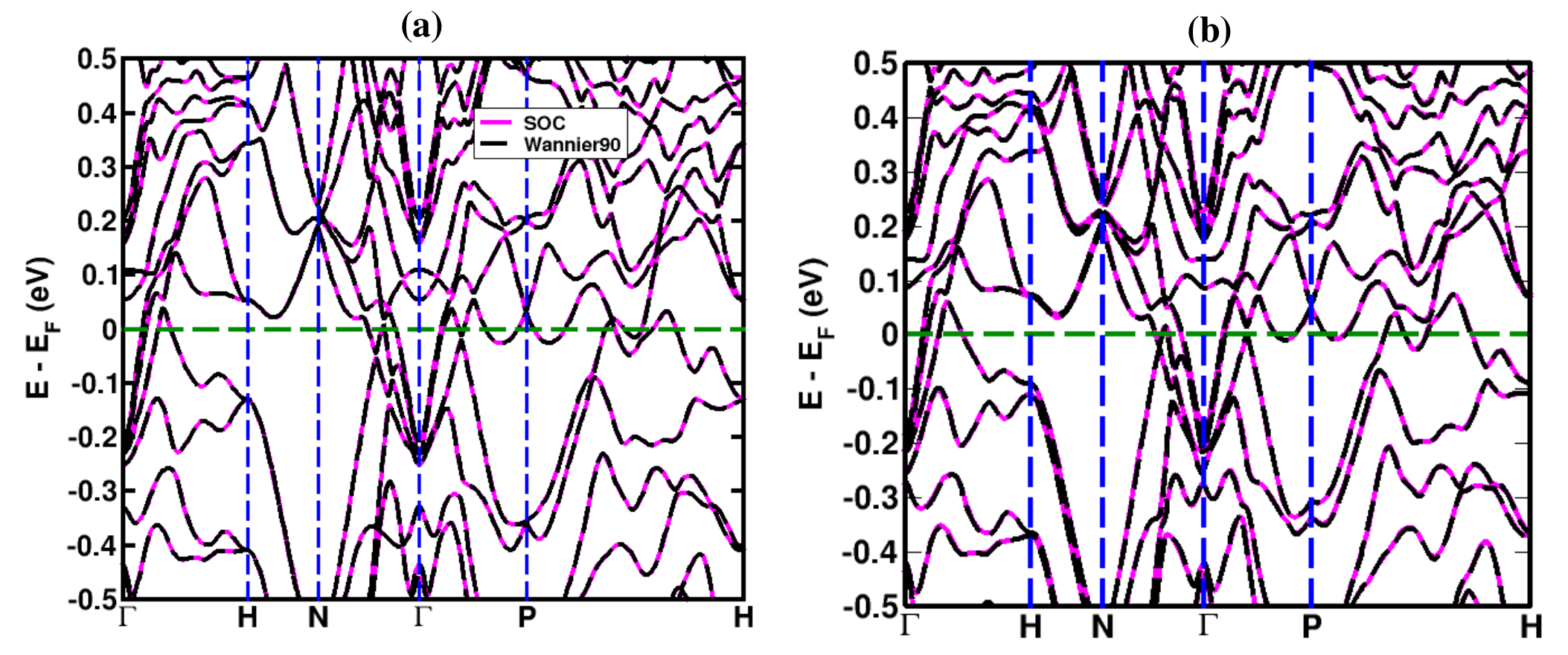}
    \caption {Electronic band structure with SOC for (a) FM and (b) Configuration-2 structures of Nd$_4$Sb$_3$. Magenta solid lines indicate the ab-initio DFT results while black-dashed lines show Wannier-interpolated bands.}
    \label{fig:band1}
\end{figure*}

\begin{figure*}[h]
    \centering
    \includegraphics[width=0.8\textwidth]{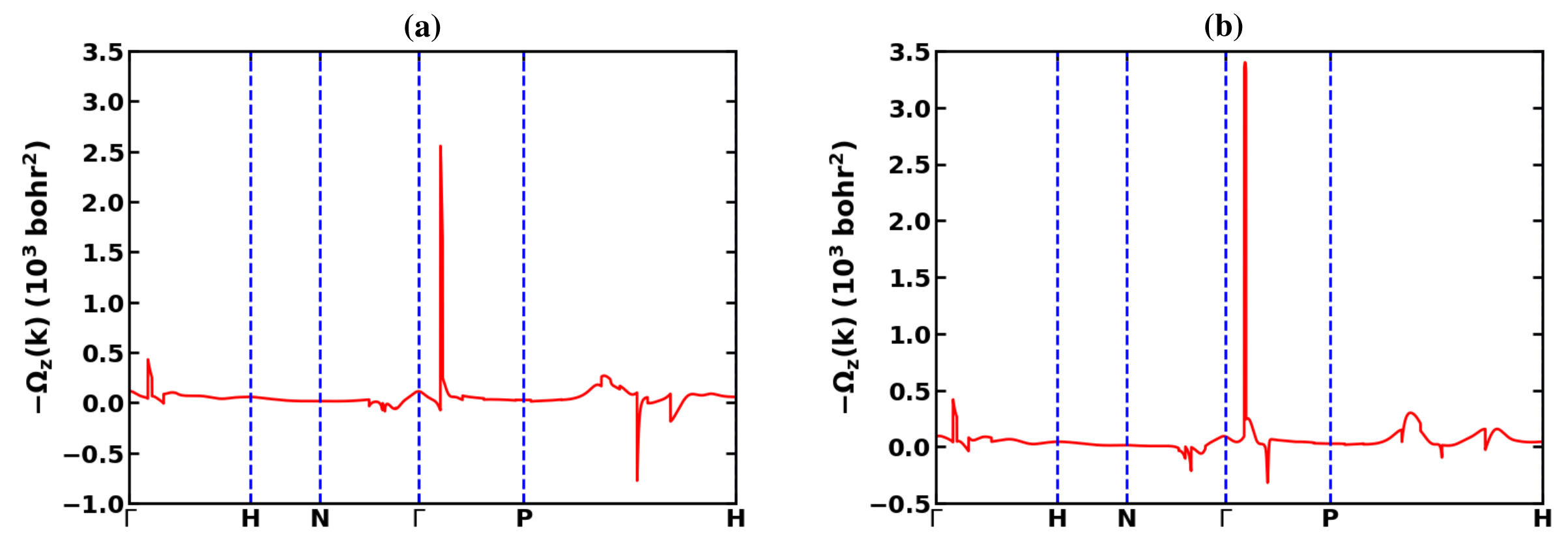}
    \caption {The calculated Berry Curvature ($\Omega_z({}\bf k)$) along the high-symmetry paths in the first Brillouin zone (BZ) for (a) FM and (b) Configuration-2 of Nd$_4$Sb$_3$.}
    \label{fig:Berry1}
\end{figure*}

\begin{table*}[h]
\centering
\caption{For FM Nd$_4$Sb$_3$, Weyl points (WPs), their Cartesian coordinates ($k_x,k_y,k_z$), relative
energy with respect to Fermi level ($\Delta\epsilon$), and chirality.}
\label{tab:weyl}

\renewcommand{\arraystretch}{1.4}

\begin{tabular}{|c|c|c|c|c|c|}
\hline

\multirow{2}{*}{\textbf{Sl. No.}}
&
\multicolumn{3}{c|}{\textbf{Cartesian coordinates of WPs}}
&
\multirow{2}{*}{\begin{tabular}{c}
\textbf{Relative enery}\\
\textbf{from E$_F$ (meV)}
\end{tabular}}
&
\multirow{2}{*}{\textbf{Chirality}}
\\

\cline{2-4}

&
\textbf{$k_x$ (\AA$^{-1}$)}
&
\textbf{$k_y$ (\AA$^{-1}$)}
&
\textbf{$k_z$ (\AA$^{-1}$)}
&
&
\\
\hline
1  &  0.126 &  0.002 & -0.152 & -64.52 & -1 \\
2  &  0.181 &  0.000 & -0.034 & -139.12 &  1 \\
3  &  0.194 &  0.014 &  0.012 & -134.02 & -1 \\
4  &  0.238 & -0.001 &  0.192 & -72.24 &  1 \\
5  & -0.238 &  0.000 &  0.191 & -71.63 &  1 \\
6  & -0.194 & -0.012 &  0.011 & -134.64 & -1 \\
7  & -0.126 & -0.001 & -0.153 & -63.34 & -1 \\
8  & -0.180 &  0.001 & -0.035 & -138.69 &  1 \\
9  & -0.107 &  0.000 &  0.002 & -199.68 &  1 \\
10 &  0.001 & -0.125 &  0.154 & -63.24 &  1 \\
11 &  0.000 &  0.125 &  0.154 & -62.95 &  1 \\
12 & -0.001 &  0.108 & -0.002 & -199.11 & -1 \\
13 & -0.194 &  0.013 &  0.012 & -134.31 & -1 \\
14 &  0.001 &  0.180 &  0.034 & -138.98 & -1 \\
15 & -0.013 &  0.194 & -0.011 & -134.46 &  1 \\
16 &  0.014 &  0.194 & -0.013 & -133.97 &  1 \\
17 &  0.334 &  0.334 &  0.000 & -130.65 &  1 \\
18 &  0.000 &  0.184 &  0.012 & -140.43 &  1 \\
19 &  0.000 &  0.238 & -0.192 & -72.18 & -1 \\
20 &  0.334 & -0.334 & -0.280 & -163.89 &  1 \\
21 & -0.334 & -0.334 & -0.281 & -164.94 &  1 \\
22 &  0.334 &  0.334 &  0.278 & -162.56 & -1 \\
23 & -0.334 &  0.334 &  0.282 & -165.39 & -1 \\
24 &  0.000 & -0.238 & -0.192 & -72.62 & -1 \\
25 & -0.011 & -0.194 & -0.011 & -134.61 &  1 \\
26 &  0.000 & -0.181 &  0.033 & -139.58 & -1 \\
27 &  0.000 & -0.108 & -0.002 & -198.86 & -1 \\
28 &  0.194 & -0.013 &  0.011 & -134.34 & -1 \\
29 &  0.195 & -0.000 &  0.007 & -135.86 &  1 \\
30 &  0.000 &  0.000 & -0.175 & -89.16 &  1 \\
31 &  0.000 & -0.000 &  0.176 & -88.95 & -1 \\
32 &  0.165 & -0.095 &  0.147 &  54.40 &  1 \\
33 &  0.107 &  0.001 &  0.260 & -13.78 &  1 \\
34 &  0.164 & -0.000 & -0.002 & -107.93 & -1 \\
35 & -0.217 & -0.141 & -0.033 &  11.49 & -1 \\
36 &  0.184 & -0.001 &  0.173 &  12.51 & -1 \\
37 &  0.274 &  0.008 & -0.074 & -32.03 &  1 \\
38 & -0.251 & -0.041 & -0.115 &   1.84 & -1 \\
39 & -0.275 & -0.009 & -0.073 & -32.07 &  1 \\
40 &  0.252 &  0.039 & -0.113 &  -0.72 & -1 \\
41 & -0.165 & -0.097 &  0.146 &  55.85 &  1 \\
42 & -0.185 &  0.003 &  0.173 &  12.11 & -1 \\
\hline
\end{tabular}
\end{table*}

\begin{table*}[t]
\centering
\caption{Weyl points (WPs), their Cartesian coordinates ($k_x,k_y,k_z$), relative energy from the Fermi level, and chirality.}
\label{tab:weyl}

\renewcommand{\arraystretch}{1.4}

\begin{tabular}{|c|c|c|c|c|c|}
\hline

\multirow{2}{*}{\textbf{Sl. No.}}
&
\multicolumn{3}{c|}{\textbf{Cartesian coordinates of WPs}}
&
\multirow{2}{*}{\begin{tabular}{c}
\textbf{Relative energy}\\
\textbf{from E$_F$ (meV)}
\end{tabular}}
&
\multirow{2}{*}{\textbf{Chirality}}
\\

\cline{2-4}

&
\textbf{$k_x$ (\AA$^{-1}$)}
&
\textbf{$k_y$ (\AA$^{-1}$)}
&
\textbf{$k_z$ (\AA$^{-1}$)}
&
&
\\
\hline
43 &  0.000 & -0.107 & -0.260 & -13.53 & -1 \\
44 & -0.163 &  0.000 & -0.002 & -108.54 & -1 \\
45 &  0.217 &  0.141 & -0.032 &  10.76 & -1 \\
46 & -0.109 & -0.000 &  0.259 & -12.42 &  1 \\
47 & -0.001 &  0.108 & -0.259 & -12.64 & -1 \\
48 &  0.165 &  0.096 &  0.147 &  55.25 &  1 \\
49 & -0.217 &  0.141 & -0.034 &  11.52 & -1 \\
50 & -0.165 &  0.096 &  0.147 &  55.74 &  1 \\
51 &  0.096 &  0.165 & -0.148 &  55.73 & -1 \\
52 & -0.141 &  0.217 &  0.033 &  11.08 &  1 \\
53 &  0.141 &  0.217 &  0.033 &  11.04 &  1 \\
54 & -0.002 &  0.275 &  0.071 & -34.06 & -1 \\
55 & -0.186 &  0.165 & -0.219 &  97.33 &  1 \\
56 & -0.034 &  0.249 &  0.119 &   1.53 &  1 \\
57 & -0.094 &  0.167 & -0.148 &  54.55 & -1 \\
58 &  0.000 &  0.164 &  0.001 & -108.86 &  1 \\
59 & -0.001 &  0.184 & -0.175 &  13.33 &  1 \\
60 & -0.166 &  0.187 &  0.220 &  97.97 & -1 \\
61 &  0.030 &  0.249 &  0.121 &   1.47 &  1 \\
62 &  0.187 &  0.166 & -0.220 &  97.87 &  1 \\
63 &  0.004 &  0.248 &  0.127 &   1.27 & -1 \\
64 &  0.165 &  0.186 &  0.219 &  97.37 & -1 \\
65 &  0.143 & -0.215 &  0.032 &  12.14 &  1 \\
66 & -0.005 & -0.275 &  0.072 & -33.12 & -1 \\
67 & -0.142 & -0.216 &  0.033 &  11.48 &  1 \\
68 & -0.041 & -0.248 &  0.119 &   4.55 &  1 \\
69 &  0.166 & -0.187 &  0.220 &  97.97 & -1 \\
70 &  0.031 & -0.248 &  0.123 &   4.01 &  1 \\
71 &  0.095 & -0.166 & -0.147 &  54.47 & -1 \\
72 &  0.000 & -0.184 & -0.174 &  12.57 &  1 \\
73 & -0.166 & -0.186 &  0.219 &  97.99 & -1 \\
74 & -0.187 & -0.167 & -0.220 &  97.82 &  1 \\
75 & -0.095 & -0.166 & -0.148 &  54.71 & -1 \\
76 &  0.010 & -0.247 &  0.130 &   3.80 & -1 \\
77 &  0.187 & -0.166 & -0.219 &  97.61 &  1 \\
78 &  0.217 & -0.141 & -0.034 &  11.31 & -1 \\
79 & -0.000 & -0.163 &  0.002 & -109.29 &  1 \\
80 & 0.192 & 0.000 & -0.012 & -151.63 & -1 \\
81 & -0.192 & 0.000 & -0.012 & -152.09 & -1 \\
82 & 0.000 & 0.143 & -0.013 & -175.33 & -1 \\
83 & 0.000 & 0.192 & 0.012 & -152.24 & 1 \\
84 & 0.000 & -0.191 & 0.012 & -150.50 & 1 \\
\hline
\end{tabular}
\end{table*}

 \begin{figure}[t]
\centering
\includegraphics[width=0.9\textwidth]{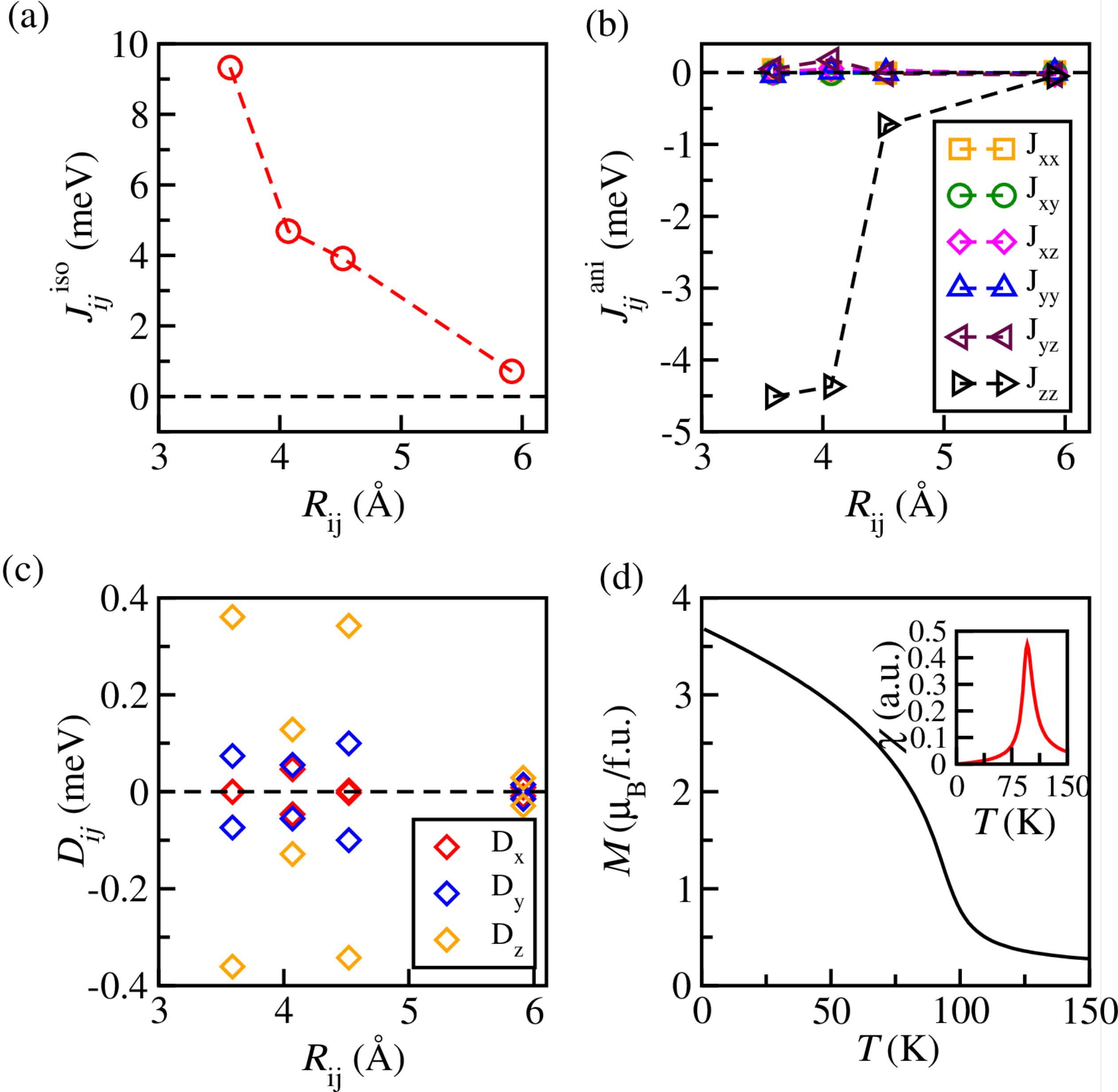}
\caption{Calculated magnetic interactions and finite-temperature magnetic response of Nd$_4$Sb$_3$ for the FM state. (a) Isotropic exchange interaction $J_{ij}^{\mathrm{iso}}$, (b) components of the symmetric anisotropic exchange tensor $J_{ij}^{\mathrm{ani}}$, and (c) components of the Dzyaloshinskii--Moriya vector $\mathbf{D}_{ij}$ as a function of the Nd--Nd interatomic distance $R_{ij}$. (d) Monte Carlo simulated magnetization $M(T)$ obtained using the exchange parameters shown in (a)-(c). The inset shows the corresponding magnetic susceptibility $\chi(T)$. The characteristic magnetic transition occurs near $\sim95$~K.}

\label{fig:col}
\end{figure}

\end{widetext}

\end{document}